\documentclass[10pt,conference]{IEEEtran}
\IEEEoverridecommandlockouts
\usepackage{cite}
\usepackage{amsmath,amssymb,amsfonts}
\usepackage{algorithmic}
\usepackage{graphicx}
\usepackage[lofdepth,lotdepth]{subfig}
\usepackage{textcomp}
\usepackage{xcolor}
\newcommand{\norm}[1]{\left\lVert#1\right\rVert}
\newtheorem{experiment}{Experiment}

\def\BibTeX{{\rm B\kern-.05em{\sc i\kern-.025em b}\kern-.08em
    T\kern-.1667em\lower.7ex\hbox{E}\kern-.125emX}}
\usepackage[ruled,vlined]{algorithm2e}   
\DontPrintSemicolon 
\usepackage{lipsum}
\usepackage{mwe}
\usepackage{subfig}
\usepackage{bbm}
\usepackage{multirow}
\usepackage[normalem]{ulem}
\usepackage{float}
\usepackage{bbm}

\usepackage{stackengine}
\def\delequal{\mathrel{\ensurestackMath{\stackon[1pt]{=}{\scriptstyle\Delta}}}}
\usepackage{comment}

\begin{document}
\pagenumbering{arabic}
\title{Unfolding for Joint Channel Estimation and Symbol Detection in MIMO Communication Systems\\
}

\author{
\IEEEauthorblockN{Swati Bhattacharya\textsuperscript{1}, K.V.S. Hari\textsuperscript{1}, Yonina C. Eldar\textsuperscript{2}}\\
\IEEEauthorblockA{\textsuperscript{1} Statistical Signal Processing Lab, Department of Electrical Communication Engineering,\\ Indian Institute of Science,
    Bangalore, India.\\
    \textsuperscript{2} Faculty of Math and Computer Science, Weizmann Institute of Science, Israel.\\
    Email: \textsuperscript{1}\{swatibh, hari\}@iisc.ac.in, \textsuperscript{2}\{yonina.eldar@weizmann.ac.il\}}\\%
}

\maketitle

\begin{abstract}

This paper proposes a Joint Channel Estimation and Symbol Detection (JED) scheme for Multiple-Input Multiple-Output (MIMO) wireless communication systems. Our proposed method for JED using Alternating Direction Method of Multipliers (JED-ADMM) and its model-based neural network version JED using Unfolded ADMM (JED-U-ADMM) markedly improve the symbol detection performance over 
JED using Alternating Minimization (JED-AM) for a range of MIMO antenna configurations. Both proposed algorithms exploit the non-smooth constraint, that occurs as a result of the Quadrature Amplitude Modulation (QAM) data symbols, to effectively improve the performance using the ADMM iterations. The proposed unfolded network JED-U-ADMM consists of a few trainable parameters and requires a small training set. We show the efficacy of the proposed methods for both uncorrelated and correlated MIMO channels. For certain configurations, the gain in SNR for a desired BER of $10^{-2}$ for the proposed JED-ADMM and JED-U-ADMM is upto $4$ dB and is also accompanied by a significant reduction in computational complexity of upto $75\%$, depending on the MIMO configuration, as compared to the complexity of JED-AM.

\end{abstract}


\section{Introduction}
Massive multiple-input multiple-output (MIMO) systems with tens or hundreds of antennas at both the transmitter and receiver have emerged as a promising technique for achieving high spectral efficiency~\cite{bjornson2016massive}. The conventional MIMO configuration has a larger number of receive antennas at the base station (BS) than the number of transmit antennas at the users (UEs). With the rise of fifth-generation wireless communication, an emerging scenario is where many users connect to a single BS with limited antennas i.e., $N \leq K$, referred to as overloaded MIMO~\cite{sah2016improved,sparse_error_recovery,ran2017sparse,IW-SOAV,TPG,dlsd,albreem2021deep}. \par
One of the critical aspects in utilizing the full potential of massive MIMO is accurate symbol detection at the BS for which Maximum Likelihood (ML) detection is optimal but computationally prohibitive. 
Linear detectors such as Zero Forcing (ZF), Minimum Mean Square Error (MMSE) or more advanced nonlinear detectors \cite{jeon2015optimality,luo2010semidefinite} and deep learning based detectors~\cite{balatsoukas2019deep,DetNet_SPAWC} perform well when $N$ is much larger than $K$ but quickly deteriorate in performance as the ratio $N/K$ approaches $1$ which is the overloaded MIMO regime. Symbol detection for overloaded MIMO has been studied in the literature using sparsity~\cite{sah2016improved,sparse_error_recovery,ran2017sparse}, convex optimization~\cite{IW-SOAV} and deep learning~\cite{TPG,dlsd,albreem2021deep,oampnet2}.

\begin{figure}[t]
    \centering
    \includegraphics[width = 0.8\linewidth]{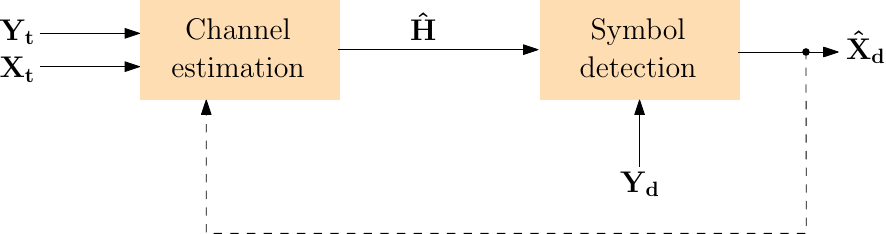}
    \caption{\footnotesize A generic JED architecture with channel estimation and symbol detection blocks that iteratively exchange information until convergence.}
    \label{fig:blockDiag}
\end{figure}

Traditionally, the tasks of channel estimation and symbol detection are decoupled and performed separately using two independent modules in cascade, in which Channel State Information (CSI) is first estimated followed by symbol detection. Joint Channel Estimation and Symbol Detection (JED) methods~\cite{blind_EP,Pbigamp,SBCE_2020,yan2019semi,Asilomar2019} consider the transmitted data as additional pilots, and exchange information iteratively between the two aforesaid modules as depicted in Fig. \ref{fig:blockDiag}, thereby resulting in better channel and symbol estimates. Message passing based techniques for JED were proposed in~\cite{blind_EP,Pbigamp,SBCE_2020,yan2019semi}, however, they depend heavily on the channel sparsity and data statistics and require extensive probability calculations. In contrast, a simple iterative technique for JED presented in \cite{Asilomar2019}, which is based on Alternating Minimization (AM), updates the channel and symbols using MMSE-based and ZF estimates, respectively. We refer to the technique proposed in \cite{Asilomar2019} as JED-AM and provide further details in Section \ref{prev_work_JEDAM}.
\par 

Deep Unfolding, a model-based neural network providing interpretability with reduced number of tunable parameters and training, was introduced in the seminal work of Gregor and LeCun in \cite{lecun_unfolding} and has gained attention over the last few years. These neural networks which are derived from a parent iterative algorithm are shown to improve the performance over fully-connected black-box detectors along with reduced computational complexity~\cite{dublid_journal,yang2020admm,usrnet,unrolling_review,DetNet_SPAWC,dlsd,TPG,albreem2021deep,balatsoukas2019deep,softoutput_haochuan,mmnet,oampnet,oampnet2,lordnet,jammer_studer,jed_cellfree,unfolding_chanest_beamform,unfolding_mixedadc,admm_hnet} and have been widely used in the domain of image processing \cite{dublid_journal,yang2020admm,usrnet,unrolling_review}.

In the context of wireless communications, unfolding has been used for symbol detection assuming knowledge of CSI, ~\cite{DetNet_SPAWC,dlsd,TPG,albreem2021deep,balatsoukas2019deep,mmnet,oampnet,lordnet} and
for jointly obtaining the channel and symbol estimates \cite{oampnet2,softoutput_haochuan,parna_icc,GEM_tsp2022}. One of the first works presented a deep unfolded version of Orthogonal AMP (OAMP), named OAMP-Net2, in~\cite{oampnet2}. An unfolded soft-output based symbol detection method is outlined in~\cite{softoutput_haochuan}. A more recent work in \cite{GEM_tsp2022} base their unfolded network on the Expectation-Maximization (EM) algorithm. A related work on JED is presented in \cite{parna_icc} which uses a convolutional neural network instead of a model-based unfolded network.

Our main contributions are summarized as follows:  
\begin{itemize}
    \item The first algorithm, 
    JED-ADMM, exploits the property of the data symbols being in the Quadrature Amplitude Modulation (QAM) constellation to impose a non-smooth constraint and propose an equivalent ADMM formulation for JED. We see from the simulation studies of Section \ref{sim_nonunfolded} that JED-ADMM results in an improved bit error rate (BER) performance over the existing JED-AM~\cite{Asilomar2019}. 
    The performance of ADMM-based algorithms greatly depends on the penalty parameter~\cite{boyd2011distributed}. We see that the BER yielded by JED-ADMM is also affected by the ADMM penalty parameter, which is decided empirically. To mitigate the loss in performance that could possibly be caused by an improper choice of the ADMM penalty parameter, we let it be trained from the data leading to the next algorithm.
    \item The second algorithm unfolds the ADMM iterations of JED-ADMM to build a model-based network (JED-U-ADMM). We also introduce some additional trainable parameters, otherwise not included in JED-ADMM, to construct a flexible unfolded network. Simulation studies in Section \ref{sim_unfolded} demonstrate the improvement in performance as well as reduction in computational complexity.
    \item We note that the previous works on JED, that use unfolding, demonstrate the performance of their algorithms for either $N>K$ in \cite{softoutput_haochuan}, $N=K$ in \cite{oampnet2} or $N \geq K$ as in \cite{GEM_tsp2022}. In contrast, we provide simulation studies for JED-ADMM and JED-U-ADMM in all possible MIMO configurations $N>K$ and $N \leq K$.
    
\end{itemize}

\par We describe the system model and existing JED in Sections \ref{system_model} and \ref{prev_work_JEDAM} respectively. The reformulation via ADMM is in Section \ref{proposed_approach} and its deep unfolded version in Section \ref{unfolded_JED_ADMM}. Simulation parameters for our experiments are described in Section \ref{simulation} and the experimental results for our proposed approach are detailed in Section \ref{experiments}.

\begin{table*}[h]
  \resizebox{\linewidth}{!}{
\begin{tabular}{|l|l|l|l|}
\hline
\multicolumn{1}{|c|}{\textbf{Symbol}} & \multicolumn{1}{c|}{\textbf{Definition}} & \multicolumn{1}{c|}{\textbf{Symbol}} & \multicolumn{1}{c|}{\textbf{Definition}} \\ \hline
$L $                                    & Number of iterations/layers in the unfolded network                            & $\mathbb{S}_\beta$                                   & rectangular $\beta$-QAM constellation                           \\ \hline
$N$                                     & Number of BS antennas  & $K$                                   & Number of single-antenna UEs       \\ \hline
$T_t$                                     & Number of pilot symbols  & $T_d$                                   & Number of data symbols      \\ \hline

$\mathbf{H}$                                     & channel between all the $K$ UEs and the BS with $N$ antennas  & $\mathbf{X}_d$                                   & data matrix transmitted from $K$ UEs over $T_d$ time instants     \\ \hline 

$\mathbf{X}_t$                                     & pilot matrix transmitted from $K$ UEs over $T_t$ time instants   & $\mathbf{V}$                                   & noise matrix at the receiver     \\ \hline

$\mathbf{Y}_t$                                     & pilot matrix received at BS over $T_t$ time instants  & $\mathbf{Y}_d$                                   & data matrix received at BS over $T_d$ time instants      \\ \hline
$\mathcal{C}$                                     & convex hull of $\mathbb{S}_\beta$  & $\mathbf{Z}_d$                                   & auxiliary variable for $\mathbf{X}_d$      \\ \hline
$\rho_c$                                     & Spatial correlation coefficient between antennas.  &      $\mathbf{\Lambda}$                           & dual variable for ADMM updates     \\ \hline
$\rho$ / $\rho_l$                                     & ADMM penalty parameter for MMSE-like update of $\tilde{\mathbf{X}}^{(l)}_d$  &      $\theta$ / $\theta_l$                           & relaxation parameter in $\tanh(.)$      \\ \hline
$\alpha$ / $\alpha_l$ & relaxation parameter for dual variable $\mathbf{\Lambda}$ & $\gamma$ / $\gamma_l$ & relaxation parameter for MMSE-like update of $\tilde{\mathbf{H}}^{(l)}$
\\ \hline
\end{tabular}}
\caption{List of Symbols}
\label{tab:list_symbols}
\end{table*}

\begin{table*}[h]
  \resizebox{\linewidth}{!}{
\begin{tabular}{|l|l|l|l|}
\hline
\multicolumn{1}{|c|}{\textbf{Symbol}} & \multicolumn{1}{c|}{\textbf{Definition}} & \multicolumn{1}{c|}{\textbf{Symbol}} & \multicolumn{1}{c|}{\textbf{Definition}} \\ \hline
JED                                    & Joint Channel Estimation and Symbol Detection                            & AM                                   & Alternating Minimization                           \\ \hline
ADMM                                     & Alternating Direction Method of Multipliers  & DL                                   & Deep Learning based       \\ \hline
CSI                                    & Channel State Information  & QAM                                   & Quadrature Amplitude Modulation      \\ \hline

\end{tabular}}
\caption{List of Acronyms}
\label{tab:list_acronyms}
\end{table*}

\par
Throughout the paper we use the following notation: Bold-faced uppercase and lowercase letters $\mathbf{A}, \mathbf{a}$ and plain faced letters $a$ represent a matrix, column vector and scalar respectively. The conjugate-transpose and the Moore-Penrose pseudoinverse of $\mathbf{A}$ are given by $\mathbf{A}^*, \mathbf{A}^\dagger$ respectively. The symbol $\mathbb{S}_\beta$ is the constellation for $\beta$-QAM and $\mathbf{I}_n$ is the $n \times n$ identity matrix. The normal distribution with mean $\mu$ and variance $\sigma^2$ is given by $x \sim \mathcal{CN}\left( \mu, \sigma^2 \right)$ and $x \sim \mathcal{U}\left[a,b\right]$ is the continuous uniform distribution on $\left[a,b\right]$. We use $\mathcal{P}_{\infty, \sqrt\beta}\left( \mathbf{X} \right)$ to denote the element-wise projection of $\mathbf{X}$ onto the infinity-norm ball of radius $\sqrt\beta$ and $\mathbbm{1}$ is the indicator function. The list of symbols and acronyms used in the paper are summarized in Tables \ref{tab:list_symbols} and \ref{tab:list_acronyms}, respectively.

\section{System Model and Problem Formulation}\label{system_model}
\begin{figure}[h]
    \centering
    \includegraphics[width = 0.8\linewidth]{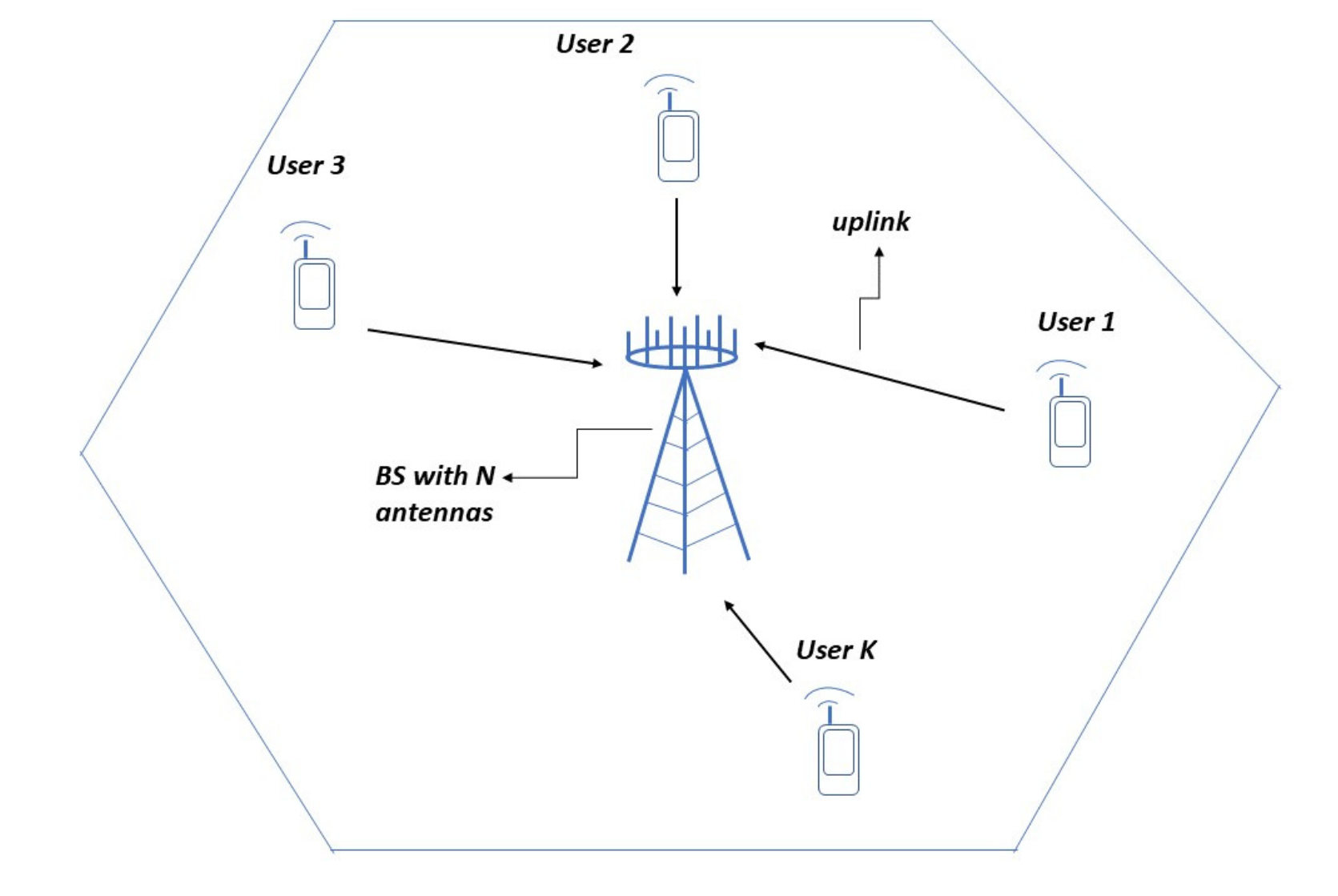}
    \caption{\footnotesize Uplink MIMO communication scenario with $N$ antennas at the BS and $K$ single-antenna UEs. }
    \label{fig:MIMOscene}
\end{figure}

Consider the uplink transmission in a narrow-band MIMO system with $N$ BS antennas and $K$ single-antenna users, as depicted in Fig. \ref{fig:MIMOscene}. 
The channel is considered to have a block fading structure and remains constant over $T$ transmissions. The first $T_t$ symbols are pilot symbols used to obtain an initial estimate of CSI and the remaining $T_d = T - T_t$ symbols are for data i.e., $\mathbf{X}_t \in \mathbb{C}^{K \times T_t}$ is the transmitted pilot matrix and $\mathbf{X}_d  \in \mathbb{C}^{K \times T_d}$ is the transmitted data matrix. At the $n^{th}$ instant, user $k$ transmits data $\{ x_k\left( n \right) \in \mathbb{S}_{\beta} : n \in \{ 1,2,..., T_d \} \}$. Here, $\mathbb{S}_{\beta} $ is the rectangular $\beta$-QAM constellation with real and imaginary parts in $\{ -\sqrt\beta + 1, -\sqrt\beta + 3,..., \sqrt\beta - 3, \sqrt\beta - 1 \}$. We consider two channel models:- (a) a Rayleigh fading channel $\mathbf{H} \in \mathbb{C}^{N \times K}$ with elements drawn from the Gaussian distribution $\mathbf{H} \sim \mathcal{CN}\left( \boldsymbol{0}, \sigma_h^2 \mathbf{I}_K \right)$ and (b) a correlated channel $\mathbf{H} = \mathbf{R}^{\frac{1}{2}}_r \mathbf{H}_w \mathbf{R}^{\frac{1}{2}}_t$, with $\mathbf{R}_r$ and $\mathbf{R}_t$ being the correlation matrices at the receiver and the transmitter respectively and $\mathbf{H}_w$ being a Rayleigh fading channel. Let $\mathbf{V} \in \mathbb{C}^{N \times T}$ be circularly symmetric complex white Gaussian noise at the receiver; $\mathbf{V} \sim \mathcal{CN}\left( \boldsymbol{0}, \sigma_v^2 \mathbf{I}_N \right)$. Denoting the received pilot and data matrix by $\mathbf{Y}_t \in \mathbb{C}^{N \times T_t}$ and $\mathbf{Y}_d  \in \mathbb{C}^{N \times T_d}$ respectively, the system model is as follows:

\begin{equation}
    \mathbf{Y} =  \mathbf{H}\mathbf{X} + \mathbf{V} \, = \, 
    \mathbf{H}\left[\mathbf{X}_t \hspace{0.2cm}  \mathbf{X}_d\right]  + \left[\mathbf{V}_t \hspace{0.2cm}  \mathbf{V}_d\right] \,=\, 
     \left[\mathbf{Y}_t \hspace{0.2cm}  \mathbf{Y}_d\right] 
\label{eq: systemModel}
\end{equation}
where $\mathbf{X} \in \mathbb{C}^{K \times T}$ is the complete transmitted signal matrix and $\mathbf{Y} \in \mathbb{C}^{N \times T}$ is the complete received signal matrix over $T$ transmissions.


\section{Background}\label{prev_work}

\subsection{Symbol Detection using ADMM}\label{prev_work_ADMM}
Symbol detection, assuming perfect CSI, using ADMM has previously been studied in \cite{ADMIN_Studer} where the idea of introducing an auxiliary variable for solving a non-smooth convex problem was shown to enhance the performance.

The task of estimating symbols $\hat{\mathbf{x}}$ from $\mathbf{y} = \mathbf{Hx} + \mathbf{v}$, with $\mathbf{y} \in \mathbb{C}^N, \mathbf{H} \in \mathbb{C}^{N \times K}$ known and $\mathbf{v}\in \mathbb{C}^N$ being the noise can be posed as \eqref{eq: actual_eq_ADMIN}. 
\begin{equation}\label{eq: actual_eq_ADMIN}
    \hat{\mathbf{x}} = \underset{\mathbf{x} \in \mathbb{S}^K_\beta}{\arg \min} \frac{1}{2} \norm{\mathbf{y} - \mathbf{Hx}}^2_2.
\end{equation}
The data symbol $x_k$, the $k^{th}$ element of $\mathbf{x}$, is  discrete valued with the probability mass function (pmf) : $\mathbb{P} \{ x_k \left( n \right) = a \} = \frac{\textstyle{1}}{\textstyle{\beta}}, \hspace{0.3cm} a \in \mathbb{S}_\beta$.
A common relaxation is to assume that the symbols are continuous variables that are uniformly distributed over $\mathcal{C}$ which is the convex hull of $\mathbb{S}_{\beta}$~\cite{Asilomar2019} i.e., $Re\{ x_k\left(n \right) \}, Im\{ x_k\left(n \right) \} \sim \mathcal{U}\left[ -\sqrt{\beta} +1, \sqrt\beta -1 \right]$,

and 
\begin{equation}\label{eq: constraint_set_C}
    \mathcal{C} = \biggl\{ \mathbf{x} \bigg|  \norm{\begin{bmatrix}
          vec\left( Re\{\mathbf{x}\} \right) \\
          vec\left( Im\{\mathbf{x}\} \right) 
         \end{bmatrix}}_\infty \leq \sqrt{\beta} - 1 \biggr\}.
\end{equation}
We can approximate the constraint $\mathbf{x} \in \mathbb{S}^K_\beta$ being in QAM constellation by its convex polytope $\mathcal{C}$ and hence, we can get the symbol estimates by solving \eqref{eq: actual_eq_ADMIN_convexpoly}.
\begin{equation}\label{eq: actual_eq_ADMIN_convexpoly}
    \hat{\mathbf{x}} = \underset{\mathbf{x} \in \mathcal{C}}{\arg \min} \frac{1}{2} \norm{\mathbf{y} - \mathbf{Hx}}^2_2.
\end{equation}

Solving \eqref{eq: actual_eq_ADMIN_convexpoly} using conventional interior-point method would incur huge complexity. Since the constraint set $\mathcal{C}$ is convex and non-smooth, the authors in \cite{ADMIN_Studer} apply the ADMM framework and rewrite the symbol detection problem as \eqref{eq: prop_eq_ADMIN}. 
\begin{equation}\label{eq: prop_eq_ADMIN}
    \hat{\mathbf{x}} = \underset{\mathbf{x}, \mathbf{z} \in \mathcal{C}}{\arg \min} \frac{1}{2} \norm{\mathbf{y} - \mathbf{Hx}}^2_2 + g\left( \mathbf{z} \right) \hspace{0.2cm} \text{s.t.} \hspace{0.2cm} \mathbf{z} = \mathbf{x},
\end{equation}
where $g\left( \mathbf{z} \right)$ is the indicator function on set $\mathcal{C}$ i.e. $\mathbbm{1}_{\{ \mathbf{z} \in \mathcal{C}\}}$. The augmented Lagrangian, in scaled form, for \eqref{eq: prop_eq_ADMIN} is 
\begin{equation}\label{eq: lagrangian_ADMIN}
    \mathcal{L}_\rho \left( \mathbf{x}, \mathbf{z}, \mathbf{\lambda} \right) = \frac{1}{2} \norm{\mathbf{y} - \mathbf{Hx}}^2_2 + \mathbbm{1}_{\{ \mathbf{z} \in \mathcal{C}\}} + \frac{\rho}{2}\norm{\mathbf{z} - \mathbf{x} - \boldsymbol{\lambda}}^2_2,
\end{equation}
with $\boldsymbol{\lambda}$ being the scaled dual variable associated with the equality constraint $\mathbf{z} = \mathbf{x}$. Simulation results in \cite{ADMIN_Studer} demonstrate that the ADMM based solution (ADMIN) outperforms linear detectors such as MMSE-based solutions in terms of the achieved BER.

\subsection{Joint Channel Estimation and Symbol Detection approaches}\label{prev_work_JEDAM}
The goal of JED is to simultaneously obtain the channel and symbol estimates $\hat{\mathbf{H}}$, $\hat{\mathbf{X}}_d$ from the known received signal $\mathbf{Y}$ and the short pilot sequence $\mathbf{X}_t$. The authors in \cite{Asilomar2019} formulate the simultaneous estimation of $\hat{\mathbf{H}}$ and $\hat{\mathbf{X}}_d$ as 
\begin{equation}\label{eq: opt_MAP-JED}
   \begin{split}
        \{ \hat{\mathbf{H}}, \hat{\mathbf{X}}_d \} =  \underset{\mathbf{H} \in \mathbb{C}^{N \times K}, \mathbf{X}_d \in \mathbb{S}_{\beta}^{K \times T_d}}{\arg \min} & \norm{\mathbf{Y}_t - \mathbf{H}\mathbf{X}_t}^2_F \\ + \norm{\mathbf{Y}_d - \mathbf{H}\mathbf{X}_d}^2_F & + \frac{\sigma_v^2}{\sigma_h^2}\norm{\mathbf{H}}^2 _F.
   \end{split}
\end{equation}
Solving \eqref{eq: opt_MAP-JED} requires a combinatorial approach, due to the discrete set $\mathbb{S}_{\beta}^{K \times T_d}$. It is common to instead solve \eqref{eq: relaxed_opt_MAP-JED} by using the convex constraint set $\mathcal{C}$ in \eqref{eq: constraint_set_C}. 
Accordingly, the JED optimization problem is given by:
\begin{equation}\label{eq: relaxed_opt_MAP-JED}
  \begin{split}
         \{ \hat{\mathbf{H}}, \hat{\mathbf{X}}_d \} = & \underset{\mathbf{H}, \mathbf{X}_d}{\arg \min}  \norm{\mathbf{Y}_t - \mathbf{H}\mathbf{X}_t}^2_F +  \norm{\mathbf{Y}_d - \mathbf{H}\mathbf{X}_d}^2_F \\ & + \frac{\sigma_v^2}{\sigma_h^2}\norm{\mathbf{H}}^2_F \hspace{0.3cm} s.t. \hspace{0.3cm} \mathbf{X}_d \in \mathcal{C}.
  \end{split}
\end{equation}
The authors of~\cite{Asilomar2019} use AM to solve \eqref{eq: relaxed_opt_MAP-JED} in an iterative manner with ZF for symbol estimates $\hat{\mathbf{X}}_d$ and MMSE for channel $\hat{\mathbf{H}}$, as summarized in Steps $1$ and $2$ of Algorithm \ref{alg: JED-AM}. 

\begin{algorithm}
\small{
\begin{algorithmic}
    \STATE \SetKwInOut{Input}{input}\SetKwInOut{Output}{output}
    \STATE \Input{$\mathbf{Y}$, $\mathbf{X}_t$}\Output{$\hat{\mathbf{X}}_d, \hat{\mathbf{H}}$}
    
    \STATE Initialize $\tilde{\mathbf{H}}^{(0)} \leftarrow \mathbf{Y}_t\mathbf{X}^*_t\left( \mathbf{X}_t\mathbf{X}^*_t + \frac{\sigma_v^2}{\sigma_h^2}\mathbf{I}_K\right)^{-1}$.
    
    \FOR {$l=1,2,\ldots,L$ }
        \STATE Step $\mathbf{1}$: $\tilde{\mathbf{X}}^{(l)}_d \leftarrow \mathcal{P}_{\infty,\sqrt \beta}\left( \tilde{\mathbf{H}}^{(l-1)^\dagger}\mathbf{Y}_d \right) $.
       
        \STATE Step $\mathbf{2}$: $\tilde{\mathbf{H}}^{(l)} \leftarrow \left(  \mathbf{Y}_t\mathbf{X}^*_t + \mathbf{Y}_d\tilde{\mathbf{X}}^{(l)^*}_d \right) \left( \mathbf{X}_t\mathbf{X}^*_t + \tilde{\mathbf{X}}^{(l)}_d\tilde{\mathbf{X}}^{(l)^*}_d + \frac{\sigma_v^2}{\sigma_h^2}\mathbf{I}_K\right)^{-1} $.
    \ENDFOR
    \STATE Final symbol and channel estimates : $\hat{\mathbf{X}}_d = \mathbb{S}_\beta \left( \tilde{\mathbf{X}}^{(L)}_d \right), \hat{\mathbf{H}} = \tilde{\mathbf{H}}^{(L)} $.
\end{algorithmic}
\caption{JED-AM~\cite{Asilomar2019}}
\label{alg: JED-AM}
}
\end{algorithm}

In the next section, we propose  our algorithms JED-ADMM and JED-U-ADMM.

\section{ADMM-based techniques for Joint Channel Estimation and Symbol Detection} 

\subsection{JED-ADMM}\label{proposed_approach}

We observe from \eqref{eq: prop_eq_ADMIN} that the optimization problem to obtain the symbol estimates consists of a smooth term $\left( \frac{\textstyle{1}}{\textstyle{2}} \norm{\mathbf{y} - \mathbf{H}\mathbf{x}}^2_2 \right)$ and a non-smooth term $\left( \mathbbm{1}_{\{ \mathbf{z} \in \mathcal{C}\}} \right)$. ADMM serves as a powerful tool in such cases~\cite{boyd2011distributed} by separating out the non-smooth part using an auxiliary variable. It has been observed that auxiliary variables provide better performance than existing methods without an auxiliary variable, in image processing applications \cite{DUBLID, dublid_journal} as well as for symbol detection in wireless communication application~\cite{ADMIN_Studer,admm_hnet}. In the image processing problem, the improvement in performance was due to updating the solution vector along three variables (auxiliary variable, blurring kernel and latent sharp image), as a result of the additional auxiliary variable, instead of the original two variables. Thus, inspired by the technique in \cite{DUBLID,dublid_journal} and the observed superior performance of ADMM over ZF/MMSE-based detectors (with perfect CSI) in \cite{ADMIN_Studer}, we propose to use ADMM for JED that updates over three variables $\mathbf{H}, \mathbf{X}_d$ and an additional auxiliary variable, $\mathbf{Z}_d \in \mathbb{C}^{K \times T_d}$.

\par 
Reformulation of the JED optimization problem using ADMM, as presented in \cite{jed_ssp}, for the original variables $\mathbf{H}, \mathbf{X}_d$ and the auxiliary variable $\mathbf{Z}_d$ becomes:

\begin{equation}\label{eq: opt_ADMM}
   \begin{split}
        \{\hat{\mathbf{H}}, \hat{\mathbf{X}}_d,  \hat{\mathbf{Z}}_d\}   =  \underset{\mathbf{H},\mathbf{X}_d,\mathbf{Z}_d} {\arg \min}  & \norm{\mathbf{Y}_t - \mathbf{H}\mathbf{X}_t}^2_F  + \norm{\mathbf{Y}_d - \mathbf{H}\mathbf{X}_d}^2_F \\ + \frac{\sigma_v^2}{\sigma_h^2}\norm{\mathbf{H}}^2_F  + & \mathbbm{1}_{\{\mathbf{Z}_d \in \mathcal{C}\}} \hspace{0.3cm} s.t. \hspace{0.2cm} \mathbf{X}_d = \mathbf{Z}_d.
   \end{split}
 \end{equation}
The term $\mathbbm{1}_{\{\mathbf{Z}_d \in \mathcal{C}\}}$ implies that $\mathbf{Z}_d$ lies in the constraint set $\mathcal{C}$ described in \eqref{eq: constraint_set_C}. The auxiliary variable $\mathbf{Z}_d$ is introduced to separate the non-smooth part of the optimization problem i.e., $\mathbbm{1}_{\{\mathbf{Z}_d \in \mathcal{C}\}} $.

The augmented Lagrangian for the model in \eqref{eq: opt_ADMM} is
\begin{equation}\label{eq: lagrangian_notscaledADMM}
\begin{split}
    \tilde{\mathcal{L}}_\rho\left( \mathbf{X}_d, \mathbf{Z}_d, \mathbf{\Lambda^{'}}, \mathbf{H}  \right) = & \norm{\mathbf{Y}_t - \mathbf{H}\mathbf{X}_t}^2_F +  \norm{\mathbf{Y}_d - \mathbf{H}\mathbf{X}_d}^2_F \\
         + \frac{\rho}{2}\norm{\mathbf{X}_d - \mathbf{Z}_d}^2_F + & \text{tr}\left( \mathbf{\Lambda}^{'*}\left( \mathbf{X}_d - \mathbf{Z}_d \right) \right) + \frac{\sigma_v^2}{\sigma_h^2}\norm{\mathbf{H}}^2_F ,
\end{split}
\end{equation}
where $\mathbf{\Lambda^{'}}$ is the Lagrangian multiplier and $\rho > 0$ is the ADMM penalty parameter.
For simplicity, we consider the scaled ADMM~\cite{boyd2011distributed} with $\mathbf{\Lambda} \delequal \frac{\textstyle{\mathbf{\Lambda}^{'}}}{\textstyle{\rho}}$ that translates the augmented Lagrangian in \eqref{eq: lagrangian_notscaledADMM} to 
\begin{equation}\label{eq: lagrangian_ADMM}
\begin{split}
    \mathcal{L}_\rho\left( \mathbf{X}_d, \mathbf{Z}_d, \mathbf{\Lambda}, \mathbf{H}  \right) = & \norm{\mathbf{Y}_t - \mathbf{H}\mathbf{X}_t}^2_F +  \norm{\mathbf{Y}_d - \mathbf{H}\mathbf{X}_d}^2_F \\ -\frac{\rho}{2}\norm{\mathbf{\Lambda}}^2_F
         + & \frac{\rho}{2}\norm{\mathbf{X}_d - \mathbf{Z}_d + \mathbf{\Lambda}}^2_F  +  \frac{\sigma_v^2}{\sigma_h^2}\norm{\mathbf{H}}^2_F.
\end{split}
\end{equation}

We next minimize $\mathcal{L}_\rho\left( \mathbf{X}_d, \mathbf{Z}_d, \mathbf{\Lambda}, \mathbf{H}  \right)$ in  \eqref{eq: lagrangian_ADMM} over the variables successively in an iterative manner as in~\cite{jed_ssp}. 
In the $l^{th}$ iteration, we denote $\tilde{\mathbf{X}}^{(l)}_d, \tilde{\mathbf{Z}}^{(l)}_d, \tilde{\mathbf{\Lambda}}^{(l)}_d\tilde{\mathbf{H}}^{(l)}_d, $ as estimates of ${\mathbf{X}}_d,{\mathbf{Z}}_d, {\mathbf{\Lambda}}_d, {\mathbf{H}}_d$, respectively. The first step in the $l^{th}$ iteration is to estimate $\tilde{\mathbf{X}}^{(l)}_d$ by solving the optimization function using the estimates of other parameters from the $(l-1)^{th}$ iteration,
\begin{equation}
    \tilde{\mathbf{X}}^{(l)}_d  = \underset{\mathbf{X}_d \in \mathbb{C}^{K \times T_d}}{\arg \min} \hspace{0.2cm} \mathcal{L}_\rho\left( \mathbf{X}_d, \tilde{\mathbf{Z}}^{(l-1)}_d, \tilde{\mathbf{\Lambda}}^{(l-1)}, \tilde{\mathbf{H}}^{(l-1)}  \right), \label{eq: opt_Xd_ADMM}
\end{equation}
whose solution, for a given $\rho$, is given by solving for $\nabla \mathcal{L}_\rho\left( \mathbf{X}_d, \tilde{\mathbf{Z}}^{(l-1)}_d, \tilde{\mathbf{\Lambda}}^{(l-1)}, \tilde{\mathbf{H}}^{(l-1)}  \right)  = \mathbf{0}$ that yields
\begin{equation}
\begin{split}
     \tilde{\mathbf{X}}^{(l)}_d \,=\, & \left( \tilde{\mathbf{H}}^{(l-1)^*}\tilde{\mathbf{H}}^{(l-1)} + \frac{\textstyle{\rho}}{\textstyle{2}}\mathbf{I} \right)^{-1} \\ 
     & \times \left( \tilde{\mathbf{H}}^{(l-1)^*}\mathbf{Y}_d +  \frac{\textstyle{\rho}}{\textstyle{2}}\left( \tilde{\mathbf{Z}}^{(l-1)}_d - \tilde{\mathbf{\Lambda}}^{(l-1)} \right) \right).
\end{split}
\end{equation}
Next, we estimate $\tilde{\mathbf{Z}}^{(l)}_d $ by solving the following optimization function using $\tilde{\mathbf{X}}^{(l)}_d, \tilde{\mathbf{\Lambda}}^{(l-1)}, \tilde{\mathbf{H}}^{(l-1)}$,
\begin{equation}
    \tilde{\mathbf{Z}}^{(l)}_d  = \underset{\mathbf{Z}_d \in \mathcal{C}}{\arg \min} \hspace{0.2cm} \mathcal{L}_\rho\left( \tilde{\mathbf{X}}^{(l)}_d, \mathbf{Z}_d, \tilde{\mathbf{\Lambda}}^{(l-1)}, \tilde{\mathbf{H}}^{(l-1)} \right). \label{eq: opt_Zd_ADMM}
\end{equation}
The solution is given by
\begin{equation}
    \tilde{\mathbf{Z}}^{(l)}_d =  \mathcal{P}_{\infty, \sqrt\beta}\left( \tilde{\mathbf{X}}^{(l)}_d + \tilde{\mathbf{\Lambda}}^{(l-1)} \right).
\end{equation}
The next step is to obtain $\tilde{\mathbf{\Lambda}}^{(l)}$ by solving 
\begin{equation}
    \underset{\mathbf{\Lambda} \in \mathbb{C}^{K \times T_d}}{\arg \min} \hspace{0.2cm}  \mathcal{L}_\rho\left( \tilde{\mathbf{X}}^{(l)}_d, \tilde{\mathbf{Z}}^{(l)}_d, \mathbf{\Lambda}, \tilde{\mathbf{H}}^{(l-1)} \right),
\end{equation}
whose solution is given by 
\begin{equation}
    \tilde{\mathbf{\Lambda}}^{(l)} = \tilde{\mathbf{\Lambda}}^{(l-1)} + \left( \tilde{\mathbf{X}}^{(l)}_d - \tilde{\mathbf{Z}}^{(l)}_d\right).
\end{equation}
The final step is to obtain a solution of $\tilde{\mathbf{H}}^{(l)}$ from
\begin{equation}
    \underset{\mathbf{H} \in \mathbb{C}^{N \times K} }{\arg \min} \hspace{0.2cm} \mathcal{L}_\rho\left( \tilde{\mathbf{X}}^{(l)}_d, \tilde{\mathbf{Z}}^{(l)}_d, \tilde{\mathbf{\Lambda}}^{(l)}, \mathbf{H}\right), \label{eq: opt_H_ADMM} 
\end{equation}
which is the MMSE solution
\begin{equation}
\begin{split}
     \tilde{\mathbf{H}}^{(l)} = &\left(  \mathbf{Y}_t\mathbf{X}^*_t + \mathbf{Y}_d\tilde{\mathbf{X}}^{(l)^*}_d \right) \\
     & \times \left( \mathbf{X}_t\mathbf{X}^*_t + \tilde{\mathbf{X}}^{(l)}_d\tilde{\mathbf{X}}^{(l)^*}_d + \frac{\textstyle{\sigma_v^2}}{\textstyle{\sigma_h^2}}\mathbf{I}_K\right)^{-1}.
\end{split}
\end{equation}

The algorithm is summarized below with the appropriate choice of the initial value of the parameters. The final estimate is obtained after $L$ iterations. \newline

\begin{algorithm*}
\small{
\begin{algorithmic}
    \STATE \SetKwInOut{Input}{input}\SetKwInOut{Output}{output}
    \STATE \Input{$\mathbf{Y}$, $\mathbf{X}_t$}\Output{$\hat{\mathbf{X}}_d, \hat{\mathbf{H}}$}
    
    \STATE Initialize $\tilde{\mathbf{Z}}^{(0)} = \boldsymbol{0}, \tilde{\mathbf{\Lambda}}^{(0)} = \boldsymbol{0}$, $\tilde{\mathbf{H}}^{(0)} \longleftarrow \mathbf{Y}_t\mathbf{X}^*_t\left( \mathbf{X}_t\mathbf{X}^*_t + \frac{\textstyle{\sigma_v^2}}{\textstyle{\sigma_h^2}}\mathbf{I}_K\right)^{-1}$.
    
    \FOR {$l=1,2,\dots,L$ }
        \STATE Step $\mathbf{1}$: update of $\mathbf{X}_d$ : 
        $\tilde{\mathbf{X}}^{(l)}_d \longleftarrow \left( \tilde{\mathbf{H}}^{(l-1)^*}\tilde{\mathbf{H}}^{(l-1)} + \frac{\textstyle{\rho}}{\textstyle{2}}\mathbf{I} \right)^{-1} \left( \tilde{\mathbf{H}}^{(l-1)^*}\mathbf{Y}_d +  \frac{\textstyle{\rho}}{\textstyle{2}}\left( \tilde{\mathbf{Z}}^{(l-1)}_d - \tilde{\mathbf{\Lambda}}^{(l-1)} \right) \right) $.
        \STATE Step $\mathbf{2}$: update of $\mathbf{Z}_d$ : $\tilde{\mathbf{Z}}^{(l)}_d \longleftarrow \mathcal{P}_{\infty, \sqrt\beta}\left( \tilde{\mathbf{X}}^{(l)}_d + \tilde{\mathbf{\Lambda}}^{(l-1)} \right) $.
        \STATE Step $\mathbf{3}$: update of $\mathbf{\Lambda}$ : $\tilde{\mathbf{\Lambda}}^{(l)} \longleftarrow \tilde{\mathbf{\Lambda}}^{(l-1)} + \left( \tilde{\mathbf{X}}^{(l)}_d - \tilde{\mathbf{Z}}^{(l)}_d\right)$.
        \STATE Step $\mathbf{4}$: update of $\mathbf{H}$ : $\tilde{\mathbf{H}}^{(l)} \longleftarrow \left(  \mathbf{Y}_t\mathbf{X}^*_t + \mathbf{Y}_d\tilde{\mathbf{X}}^{(l)^*}_d \right) \left( \mathbf{X}_t\mathbf{X}^*_t + \tilde{\mathbf{X}}^{(l)}_d\tilde{\mathbf{X}}^{(l)^*}_d + \frac{\textstyle{\sigma_v^2}}{\textstyle{\sigma_h^2}}\mathbf{I}_K\right)^{-1} $.
    \ENDFOR
    \STATE Final symbol and channel estimates : $\hat{\mathbf{X}}_d = \mathbb{S}_\beta \left( \tilde{\mathbf{X}}^{(L)}_d \right), \hat{\mathbf{H}} = \tilde{\mathbf{H}}^{(L)} $.
\end{algorithmic}
\caption{Algorithm for Joint Channel Estimation and Symbol Detection using ADMM (JED-ADMM)}
\label{alg: JED-ADMM}
}
\end{algorithm*}

\subsection{Computational Complexity}
The computational complexity in terms of Floating Point Operations (FLOPS) for the proposed JED-ADMM and existing JED-AM is presented next. We note that the multiplication of two matrices of size $m\times n$ and $n\times p$ requires $mnp$ FLOPS and the  inversion of a square matrix of size $n$ needs $n^3$ FLOPS.
Both JED-ADMM and JED-AM use MMSE for the channel update. However, JED-ADMM needs an extra $NKT_d$ FLOPS per iteration for the symbol detection step. The summary of the complexity of the algorithms is given in Table \ref{tab:detailed_FLOPScount_withmmsezf}. As an example,  considering $N=16,K=16,T_d = 512, T_t = K$, we observe that JED-ADMM requires $557056$ FLOPS per iteration and JED-AM needs $425984$ FLOPS, the former being $\approx 1.3$ times more expensive than the latter. 
\begin{table*}[h]
\centering
\resizebox{\textwidth}{!}{
\begin{tabular}{|cccccc|}
\hline
\multicolumn{4}{|c|}{Computational complexity in terms of FLOPS for each step} \\ \hline
\multicolumn{2}{|c|}{proposed JED-ADMM in Algorithm \ref{alg: JED-ADMM}} & \multicolumn{2}{c|}{JED-AM in Algorithm \ref{alg: JED-AM} \cite{Asilomar2019}} \\ \hline
\multicolumn{1}{|l|}{initial estimate $\tilde{\mathbf{H}}^{(0)}$} & \multicolumn{1}{l|}{$NKT_t + K^2T_t + K^2N + K^3$} & \multicolumn{1}{l|}{initial estimate $\tilde{\mathbf{H}}^{(0)}$} & \multicolumn{1}{l|}{$NKT_t + K^2T_t + K^2N + K^3$}  \\ \hline
\multicolumn{1}{|c|}{Steps in $l^{th}$ iteration} & \multicolumn{1}{c|}{FLOPS} & \multicolumn{1}{c|}{Steps  in $l^{th}$ iteration} & \multicolumn{1}{c|}{FLOPS} \\ \hline
\multicolumn{1}{|c|}{Step 1 : $\tilde{\mathbf{X}}^{(l)}_d$} & \multicolumn{1}{c|}{$K^2N + K^3 + KNT_d + K^2T_d$} & \multicolumn{1}{c|}{Step 1 : $\tilde{\mathbf{X}}^{(l)}_d$} & \multicolumn{1}{c|}{$K^3 + K^2N + KNT_d + 2KT_d$} \\ \hline
\multicolumn{1}{|c|}{Step 2 : $\tilde{\mathbf{Z}}^{(l)}_d$} & \multicolumn{1}{c|}{$2KT_d$} & \multicolumn{1}{c|}{Step 2 : $\tilde{\mathbf{H}}^{(l)}_d$} & \multicolumn{1}{c|}{$NK(T_t+T_d) + K^2(T_t + T_d) + K^2N$} \\ \hline
\multicolumn{1}{|c|}{Step 3 : $\tilde{\mathbf{\Lambda}}^{(l)}_d$} & \multicolumn{1}{c|}{-} & \multicolumn{1}{c|}{} & \multicolumn{1}{c|}{-} \\ \hline
\multicolumn{1}{|c|}{Step 4 : $\tilde{\mathbf{H}}^{(l)}_d$} & \multicolumn{1}{c|}{$NK(T_t+T_d) + K^2(T_t + T_d) + K^2N$} & \multicolumn{1}{c|}{} & \multicolumn{1}{c|}{-} \\ \hline
\multicolumn{4}{|c|}{Total no. of FLOPS over $L$ iterations} \\ \hline
\multicolumn{1}{|c|}{JED-ADMM} & \multicolumn{1}{c|}{$(3K^2N + K^2(2T_d + 2T_t) + NK(2T_d + 2T_t)+ 2K^3)L$} & \multicolumn{1}{c|}{JED-AM} & \multicolumn{1}{c|}{$(3K^2N + K^2(2T_d + 2T_t) + NK(T_d + 2T_t)+ 2K^3)L$} \\ \hline
\end{tabular}
}
\caption{\footnotesize Comparison of FLOPS (multiplication) required for the baseline algorithms with $L$ denoting the number of iterations in JED-AM and JED-ADMM.}
\label{tab:detailed_FLOPScount_withmmsezf}
\end{table*}
Considering a larger MIMO system $N=64, K = 80,T_d = 512, T_t = K$, we find that JED-ADMM needs $14315520$ FLOPS, which is $ \approx 1.22$ times more than that of JED-AM. Hence, we note from the above examples that as $N$ and $K$ increase, this difference becomes negligible because the matrix inversion steps dominate  the complexity.
\newline A concise version of our work on JED-ADMM for overloaded MIMO $\left( N \leq K \right)$ systems was presented in \cite{jed_ssp}. We demonstrate, in Experiment $1$ of Section \ref{sim_nonunfolded}, that the proposed JED-ADMM algorithm outperforms JED-AM for properly chosen values of the ADMM penalty $\rho$. This variation in performance of JED-ADMM, depending on the value of $\rho$, motivates the development of JED-U-ADMM as detailed in Section \ref{unfolded_proposed_approach}.

\section{Unfolding Framework for Deep Learning based Joint Estimation and Detection}\label{unfolded_proposed_approach}

We observed from Figs. \ref{fig:JED_32X32_iid_varyrho_1}, \ref{fig:JED_32X16_iid_varyrho_1} and \ref{fig:JED_64X80_iid_varyrho_1} that the choice of $\rho$ and number of iterations are critical for JED-ADMM. To mitigate the decline in performances that could be caused by poor choice of these parameters, we propose to use deep unfolding on JED-ADMM framework by introducing trainable parameters in the update equations.

\subsection{Unfolding of JED-ADMM}\label{unfolded_JED_ADMM}
We follow the approach of \cite{oampnet2,TPG} and introduce only scalar learnable quantities to propose am unfolded network JED-U-ADMM, that is derived from the unfolding of the ADMM iterations of JED-ADMM. In Steps $1$-$4$ of JED-ADMM (Algorithm \ref{alg: JED-ADMM}) the right hand side of the equation is essentially an affine combination of two or more terms with the weighting coefficient set to $1$. To allow for flexibility in the framework, we introduce additional parameters $\theta_l,\alpha_l$ that do not necessarily restrict the weighting coefficients to $1$. Furthermore, we replace the fixed quantity $\frac{\textstyle{\sigma^2_v}}{\textstyle{\sigma^2_h}}$ with $\rho_l$ and $\gamma_l$ which can be learnt from the data. We then learn $\rho_l,\theta_l,\alpha_l,\gamma_l$ from data. \par
We note that the system model as described in \eqref{eq: systemModel} has complex-valued variables. We recast the model using real-valued variables for ease of implementation using 
PyTorch~\cite{paszke2019pytorch}, as:
\begin{equation}
    \overline{\mathbf{Y}} = \overline{\mathbf{H}}\, \overline{\mathbf{X}} + \overline{\mathbf{V}} \label{eq: real_systemModel},
\end{equation}
where
\begin{align*}
    \overline{\mathbf{Y}} = & \left[ Re\left( \mathbf{Y} \right)^T Im\left( \mathbf{Y} \right)^T \right]^T, \hspace{0.2cm} \overline{\mathbf{X}} =  \left[ Re\left( \mathbf{X} \right)^T Im\left( \mathbf{X} \right)^T \right]^T. \\
    & \overline{\mathbf{H}} =  \begin{bmatrix}
Re\left( \mathbf{H} \right) & -Im\left( \mathbf{H} \right)\\
Im\left( \mathbf{H} \right) & Re\left( \mathbf{H} \right)
\end{bmatrix}.  
\end{align*}

Similar to the pilot and data portions of $\mathbf{X}$ in \eqref{eq: systemModel}, we denote the respective real counterparts of pilot and data using the overline $\overline{(.)}$, as $\overline{\mathbf{X}}_t$, $\overline{\mathbf{X}}_d$, $\overline{\mathbf{Y}}_t$, $\overline{\mathbf{Y}}_d$. The corresponding real auxiliary variable is now denoted by $\overline{\mathbf{Z}}_d$ and $\overline{\mathbf{\Lambda}}$ denotes the real dual variable for the ADMM updates.

\par 

Solving for the Lagrangian in a similar manner as in \eqref{eq: lagrangian_ADMM} with the real-valued variables of \eqref{eq: real_systemModel}, we obtain
\begin{equation}\label{eq: real_lagrangian_ADMM}
\begin{split}
    \mathcal{L}_\rho\left( \overline{\mathbf{X}}_d, \overline{\mathbf{Z}}_d, \overline{\mathbf{\Lambda}}, \overline{\mathbf{H}}  \right) = & \norm{\overline{\mathbf{Y}}_t - \overline{\mathbf{H}} \,\overline{\mathbf{X}}_t}^2_F +  \norm{\overline{\mathbf{Y}}_d - \overline{\mathbf{H}}\,\overline{\mathbf{X}}_d}^2_F \\ -\frac{\rho}{2}\norm{\mathbf{\overline{\Lambda}}}^2_F
         + & \frac{\rho}{2}\norm{\overline{\mathbf{X}}_d - \overline{\mathbf{Z}}_d + \overline{\mathbf{\Lambda}}}^2_F  +  \frac{\sigma_v^2}{\sigma_h^2}\norm{\overline{\mathbf{H}}}^2_F.
\end{split}
\end{equation}

We next minimize $\mathcal{L}_\rho\left( \overline{\mathbf{X}}_d, \overline{\mathbf{Z}}_d, \overline{\mathbf{\Lambda}}, \overline{\mathbf{H}}  \right)$ in  \eqref{eq: real_lagrangian_ADMM} over the variables successively using a multi-layered neural network. 
In the $l^{th}$ layer, we denote $\tilde{\overline{\mathbf{X}}}^{(l)}_d, \tilde{\overline{\mathbf{Z}}}^{(l)}_d, \tilde{\overline{\mathbf{\Lambda}}}^{(l)}_d,\tilde{\overline{\mathbf{H}}}^{(l)}_d, $ as estimates obtained after that layer. 

Successive minimization over the variables for $\mathcal{L}_\rho\left( \overline{\mathbf{X}}_d, \overline{\mathbf{Z}}_d, \overline{\mathbf{\Lambda}}, \overline{\mathbf{H}}  \right)$ yield similar update equations as Steps $1$-$4$ in Algorithm \ref{alg: JED-ADMM} with $\tilde{\mathbf{X}}^{(l)}_d$, $\tilde{\mathbf{Z}}^{(l)}_d$, $\tilde{\mathbf{\Lambda}}^{(l)}$, $\tilde{\mathbf{H}}^{(l)}$ being replaced by their real counterparts $\tilde{\overline{\mathbf{X}}}^{(l)}_d$, $\tilde{\overline{\mathbf{Z}}}^{(l)}_d$, $\tilde{\overline{\mathbf{\Lambda}}}^{(l)}$, $\tilde{\overline{\mathbf{H}}}^{(l)}$, respectively.

\par
\begin{figure*}[htb]
    \centering
    \includegraphics[width = 0.78\linewidth]{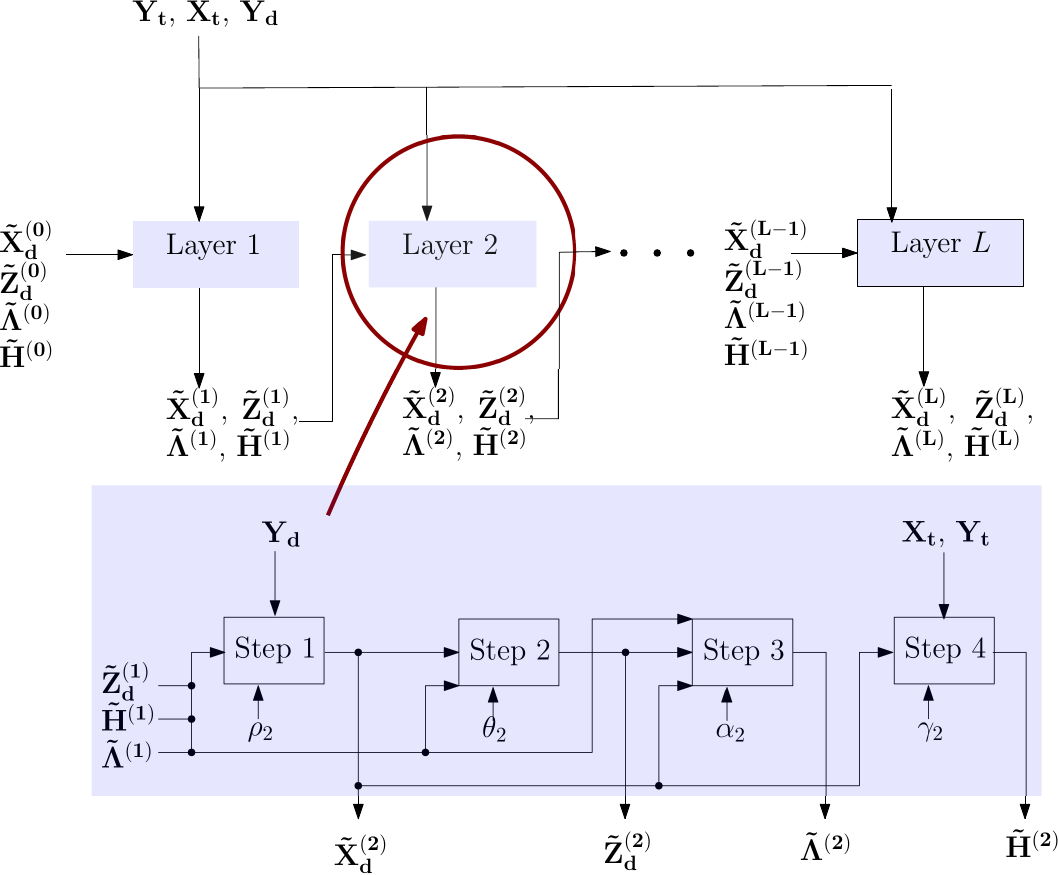}
    \caption{\footnotesize Block diagram of the JED-U-ADMM architecture. There are $L$ cascaded layers, each with $4$ trainable parameters $\rho_l, \theta_l,\alpha_l,\gamma_l$. Steps $1,2,3,4$ refer to the corresponding steps in Algorithm \ref{alg: JED-U-ADMM}.}
    \label{fig:blockDiag_DLJEDADMM}
\end{figure*}

 {\em Network architecture}: The network architecture of  JED-U-ADMM network is obtained by unfolding each of the iterations of JED-ADMM. There are $L$  cascaded layers in the network, with the same architecture, but with different trainable parameters $\Theta_l$ in each of the $l^{th}$ layer. We provide a pictorial description of the JED-U-ADMM in Fig. \ref{fig:blockDiag_DLJEDADMM}. 
\par For the purpose of unfolding, we introduce trainable parameters in each of the update equations in the following manner: 

\begin{itemize}
    \item In Step $1$ for the MMSE-like update of $\tilde{\overline{\mathbf{X}}}_d$, we replace the fixed ADMM penalty parameter $\rho$ with $\rho_l$ that can vary with the layers $l$.
    
    \item To incorporate the non-linearity of neural networks, we use the $\tanh$ activation function for the update of the auxiliary variable $\tilde{\overline{\mathbf{Z}}}_d$ in Step $2$, instead of $\mathcal{P}_{\infty, \sqrt \beta}$.
    
    \item In Step $3$, the error term $\left( \tilde{\overline{\mathbf{X}}}^{(l)}_d - \tilde{\overline{\mathbf{Z}}}^{(l)}_d\right)$ has a coefficient of $1$. We can modify it to having $\alpha_l$ as the coefficient and let the data decide the optimal value of $\alpha_l$.

    \item For the MMSE-like update of $\tilde{\overline{\mathbf{H}}}$ in Step $4$, we replace the fixed regularization parameter $\frac{\textstyle{\sigma_v^2}}{\textstyle{\sigma_h^2}}$, that is dependent on the noise variance and might not always be known, with a trainable parameter $\gamma_l$. 
\end{itemize}
We summarize the above steps for JED-U-ADMM in Algorithm \ref{alg: JED-U-ADMM}. Note that for $\rho_l, \gamma_l = \frac{\textstyle{\sigma^2_v}}{\textstyle{\sigma^2_h}}$, $\alpha_l = 1$ and by replacing the $\tanh$ function with $\mathcal{P}_{\infty, \sqrt \beta}$ (the standard projection function in literature), we get back the JED-ADMM algorithm in Algorithm \ref{alg: JED-ADMM}.

\par

{\em Training} : The loss function for backpropagation is defined as $\mathbf{L}_\Theta = \norm{\overline{\mathbf{X}}_d - \tilde{\overline{\mathbf{X}}}^{(L)}_d}^2_F$. 
\par In the training phase of JED-U-ADMM, we start with an initial value of $\Theta_l$ which is updated in accordance with the loss function $\mathbf{L}_\Theta$ in each training epoch. The update rule for $\Theta_l$ is determined by the Adam optimizer~\cite{kingma2014adam}. Once the network is trained for enough number of epochs, the weights ( i.e., $\Theta_l$ values ) are frozen and we use this trained network to detect/estimate the unknown data. 
Hence, basically the equations governing JED-U-ADMM are obtained following the framework of JED-ADMM where the parameters $\Theta_l$ are obtained in a data-driven manner. 

{\em Computational complexity :} In the training phase, since JED-U-ADMM is a neural network, there is an accompanying training cost which depends on the number of trainable parameters. For the JED-U-ADMM architecture summarized in Algorithm \ref{alg: JED-U-ADMM}, it can be seen that we have distinct trainable parameters for each of the $L$ layers leading to the training of $4L$ parameters. 

\begin{algorithm*}[tbh]
\small{
\begin{algorithmic}
    \STATE \SetKwInOut{Input}{input}\SetKwInOut{Output}{output}
    \STATE \Input{$\overline{\mathbf{Y}}$, $\overline{\mathbf{X}}_t$}\Output{$\hat{\overline{\mathbf{X}}}_d, \hat{\overline{\mathbf{H}}}$}
    \STATE Trainable parameters : $\Theta_l = \{ \rho_l, \theta_l, \alpha_l, \gamma_l\}, \hspace{0.2cm} l = 1,2,\ldots,L$.
    \STATE Initialize $\tilde{\overline{\mathbf{Z}}}_d^{(0)} = \boldsymbol{0}, \tilde{\overline{\mathbf{\Lambda}}}^{(0)} = \boldsymbol{0}$,
$\tilde{\overline{\mathbf{H}}}^{(0)} \longleftarrow \overline{\mathbf{Y}}_t\overline{\mathbf{X}}^*_t\left( \overline{\mathbf{X}}_t \overline{\mathbf{X}}^*_t + \gamma_0\mathbf{I}_K\right)^{-1}$.

    \FOR {$l=1,2,\ldots,L$ }
        \STATE Step $\mathbf{1}$: update of $\overline{\mathbf{X}}_d$ : $\tilde{\overline{\mathbf{X}}}^{(l)}_d \longleftarrow \left( \tilde{\overline{\mathbf{H}}}^{(l-1)^*}\tilde{\overline{\mathbf{H}}}^{(l-1)} + \rho_l\mathbf{I} \right)^{-1} \left( \tilde{\overline{\mathbf{H}}}^{(l-1)^*}\overline{\mathbf{Y}}_d +  \rho_l\left( \tilde{\overline{\mathbf{Z}}}^{(l-1)}_d - \tilde{\overline{\mathbf{\Lambda}}}^{(l-1)} \right) \right) $.
        \STATE Step $\mathbf{2}$: update of $\overline{\mathbf{Z}}_d$ : $\tilde{\overline{\mathbf{Z}}}^{(l)}_d \longleftarrow \tanh\left( \frac{\textstyle{\tilde{\overline{\mathbf{X}}}^{(l)}_d + \tilde{\overline{\mathbf{\Lambda}}}^{(l-1)}_d}}{\textstyle{|\theta_l|}} \right) $.
        \STATE Step $\mathbf{3}$: update of $\overline{\mathbf{\Lambda}}$ : $\tilde{\overline{\mathbf{\Lambda}}}^{(l)} \longleftarrow \tilde{\overline{\mathbf{\Lambda}}}^{(l-1)} + \alpha_l\left( \tilde{\overline{\mathbf{X}}}^{(l)}_d - \tilde{\overline{\mathbf{Z}}}^{(l)}_d\right)$.
        \STATE Step $\mathbf{4}$: update of $\overline{\mathbf{H}}$ : $\tilde{\overline{\mathbf{H}}}^{(l)} \longleftarrow \left(  \overline{\mathbf{Y}}_t\overline{\mathbf{X}}^*_t + \overline{\mathbf{Y}}_d\overline{\mathbf{X}}^{(l)^*}_d \right) \left( \overline{\mathbf{X}}_t\overline{\mathbf{X}}^*_t + \tilde{\overline{\mathbf{X}}}^{(l)}_d \overline{\mathbf{X}}^{(l)^*}_d + \gamma_l\mathbf{I}_K\right)^{-1} $.
    \ENDFOR
    \STATE Final symbol and channel estimates : $\hat{\overline{\mathbf{X}}}_d = \mathbb{S}_\beta \left( \tilde{\overline{\mathbf{X}}}^{(L)}_d \right), \hat{\overline{\mathbf{H}}} = \tilde{\overline{\mathbf{H}}}^{(L)} $.
\end{algorithmic}
\caption{Proposed Algorithm for Unfolding based Joint Channel Estimation and Symbol Detection using ADMM (JED-U-ADMM)}
\label{alg: JED-U-ADMM}
}
\end{algorithm*}

{\em Shared parameter architecture :} A variation could be to allow some of the parameters to be shared across the layers. The reasoning for this alternate unfolded architecture arises from the intuition that 
it is possible that the values of some of the parameters do not vary significantly across the layers. In such cases, parameter sharing can lead to reduction of training complexity while maintaining similar performance. Hence, we propose the following shared architecture, henceforth referred to as \lq JED-U-ADMM : shared Params'.
The parameters $\alpha_l,\theta_l, \gamma_l$ are shared across all the $L$ layers which effectively replaces these $3L$ parameters with just $3$ trainable parameters i.e. $\alpha, \theta, \gamma$. Thus, in the Steps $2,3,4$ of Algorithm \ref{alg: JED-U-ADMM}, $\theta_l = \theta$, $\alpha_l = \alpha$ and $\gamma_l = \gamma$ respectively. This reduces the total trainable parameters
to only $L + 4$ trainable parameters. Note that for the initial $\tilde{\overline{\mathbf{H}}}^{(0)}$ update, we use $\gamma_0$. 
In the estimation phase, the computational complexity of JED-ADMM and JED-U-ADMM is  the same (Table \ref{tab:detailed_FLOPScount_withmmsezf}).

In the next section we study the performance of the proposed methods for different scenarios.

\section{Simulation Results}
We study the BER performance of our proposed algorithms JED-U-ADMM, JED-ADMM and JED-AM. 
\subsection{Simulation scenario:}\label{simulation} In this paper, we consider an uplink scenario where the BS with $N$ antennas receives signals from  $K$ UEs each with a single transmit antenna. We assume the UEs are geographically well separated and the signals propagate through independent channels.

{\em Transmitter parameters:} The pilot matrix $\mathbf{X}_t$ consists of columns from the Discrete Fourier Transform (DFT) matrix and the data matrix $\mathbf{X}_d$ is sampled from a $4$-QAM $\left( \beta = 4 \right)$ constellation. Thus, the constellation set is given by $\mathbb{S}_\beta = \{ 1 + j, 1 - j, -1 + j, -1 -j \}$ and the average energy per symbol, denoted by $E_s$ equals $2$. The pilot length $T_t$ for obtaining an initial CSI was set to the optimal pilot length required i.e., $T_t = K$. 

{\em Channel:} The channel is assumed to be constant over $T = 300$ transmissions. We consider two cases of the channel:  i.i.d fading Rayleigh channel, $\mathbf{H} \sim \mathcal{CN}\left( \mathbf{0}, \frac{1}{K}\mathbf{I}\right)$ and correlated channels $\mathbf{H} = \mathbf{R}_r^{\frac{1}{2}}\mathbf{H}_w\mathbf{R}_t^{\frac{1}{2}}$, with $\mathbf{H}_w$ being an i.i.d Rayleigh fading channel. The correlation matrices are $\mathbf{R}_r$ in \eqref{eq: corrAntMat} and $\mathbf{R}_t = \mathbf{I}_K$ at the receiver and transmitter respectively~\cite{correlationMIMOchannel}. 
\begin{equation}\label{eq: corrAntMat}
    \mathbf{R}_r =  \begin{bmatrix}
1 & \rho_c & ... & \rho^{K-2}_c & \rho^{K-1}_c\\
\rho_c & 1 & \rho_c & ... & \rho^{K-2}_c\\
... & ... & ... & ... & ... \\
\rho^{K-1}_c & \rho^{K-2}_c & ... & ... & 1
\end{bmatrix}.
\end{equation}
In this paper, we consider an uplink channel model which justifies the introduction of a correlation matrix at the BS antennas at the receiver. The users are assumed to be geographically separated and hence uncorrelated. 

{\em Receiver algorithms:} For JED-ADMM, the receiver implements the iterative algorithms described in Algorithm \ref{alg: JED-ADMM} where $L$ refers to the number of iterations. $\rho$ is chosen empirically in a manner as in \cite{ADMIN_Studer}. For the unfolded JED-U-ADMM, we use Algorithm \ref{alg: JED-U-ADMM} to train the network over $100/200$ epochs to get the trained values of $\Theta_l$. In the initial $T_t$ transmissions when the transmitter sends pilots to the receiver to obtain an initial CSI, we choose a pilot structure such that only one antenna sends a pilot at a particular time instant and the rest of the antennas are turned off.  During the detection phase we assume data transmitted over all antennas and use the trained $\Theta_l$ to detect symbols. For each channel realization, $T = (512 + T_t)$ bits are sent and $2\times 10^4$ channel realizations are used. We use the PyTorch~\cite{paszke2019pytorch} environment with Adam optimizer. 

{\em Performance Metric:} The performance metric to compare the algorithms is chosen to be the bit error rate (BER). We compute the BER based on errors observed in $5 \times 10^6$ bits from each of the transmitter antennas. In all the experiments, we define the Signal-to-Noise Ratio ($\text{$\text{SNR}$}$) per receiver antenna as 
    \begin{equation}\label{eq: snrdef_journal}
        \text{$\text{SNR}$} = \frac{\mathbf{E}\left[ \norm{\mathbf{H}\mathbf{X}_d}^2 \right]}{\mathbf{E}\left[ \norm{\mathbf{V}_d}^2 \right]} = \frac{E_s}{\sigma^2_v}.
    \end{equation}

\subsection{Experiments}\label{experiments}
We consider experiments to study the performance of the proposed JED-U-ADMM, JED-ADMM and JED-AM~\cite{Asilomar2019} methods.

\subsubsection{Study the performance of JED-AM and JED-ADMM}\label{sim_nonunfolded}
We first study the performance of JED-ADMM and JED-AM under various MIMO configurations and channel conditions.

\begin{experiment}{\textit{Study the effect of varying $\rho$}}\\
In this experiment, we study how the choice of the ADMM penalty parameter $\rho$ affects the BER performance of JED-ADMM. Three different MIMO configurations are considered - $N > K$, $N=K$ and $N < K$. For $N<=K$, $20$ iterations of JED-ADMM is considered while $100$ iterations were taken for $N>K$.  We consider two values of $\rho = \frac{\textstyle{\sigma^2_v}}{\textstyle{\sigma^2_h}}$ and $\rho = 4 \times \frac{\textstyle{\sigma^2_v}}{\textstyle{\sigma^2_h}}$.

\begin{figure}[h]
    \centering
    \includegraphics[width = 0.8\linewidth]{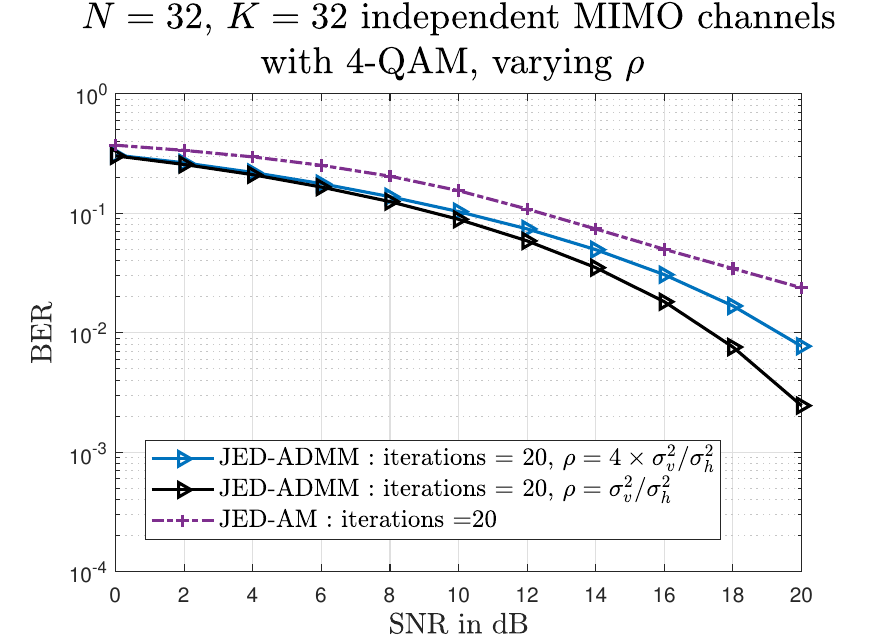}
    \caption{ \footnotesize $N=K$ : This plot shows the variation of BER vs SNR as $\rho$ is varied for $20$ iterations of JED-ADMM and JED-AM in a $32 \times 32$ MIMO for independent Rayleigh channels.}
    \label{fig:JED_32X32_iid_varyrho_1}
\end{figure}

\begin{figure}[h]
    \centering
    \includegraphics[width = 0.8\linewidth]{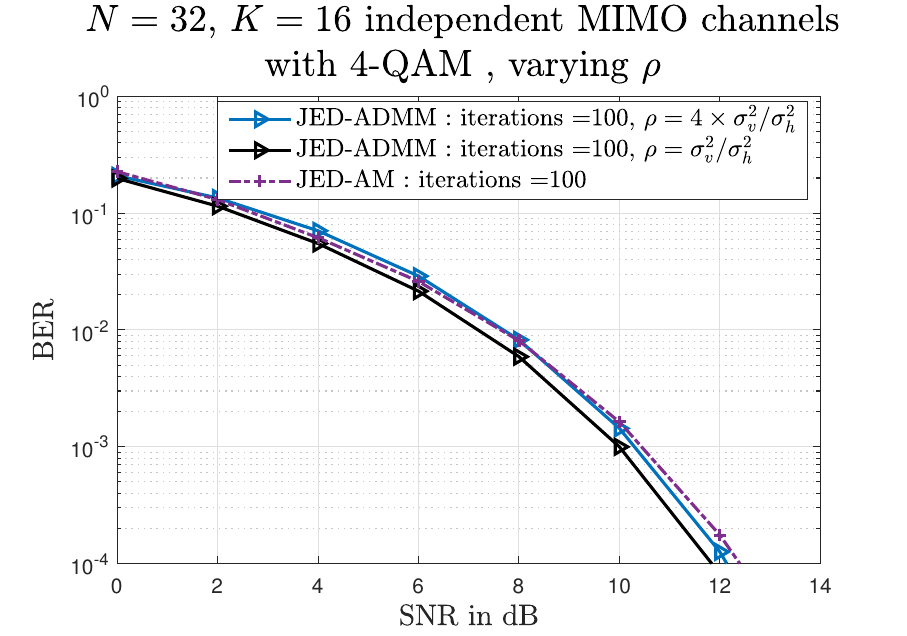}
    \caption{ \footnotesize $N>K$ : This plot shows the variation of BER vs SNR as $\rho$ is varied for $100$ iterations of JED-ADMM and JED-AM in a $32 \times 16$ MIMO for independent Rayleigh channels.}
    \label{fig:JED_32X16_iid_varyrho_1}
\end{figure}

\begin{figure}[h]
    \centering
    \includegraphics[width = 0.8\linewidth]{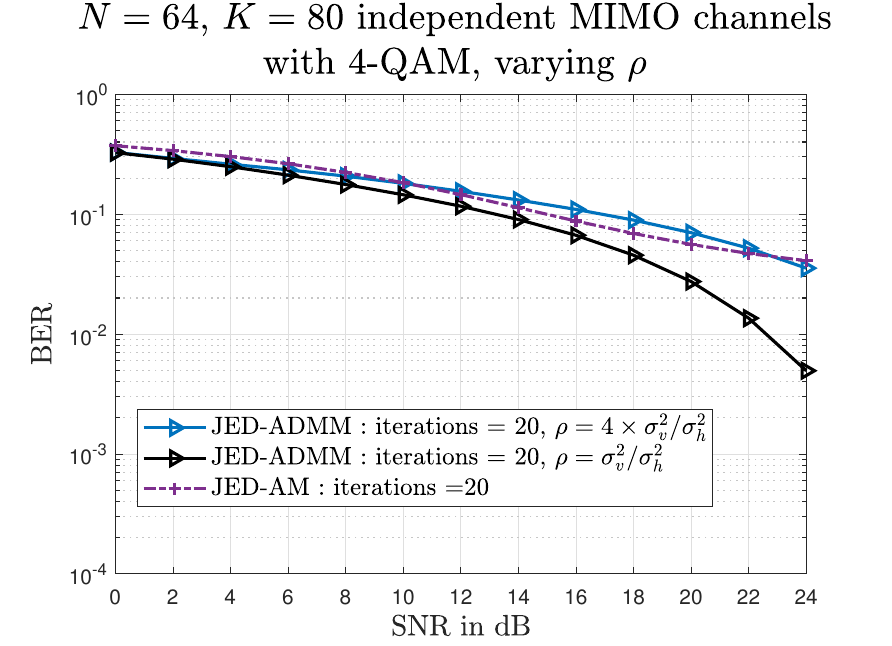}
    \caption{ \footnotesize $N < K$ : This plot shows the variation of BER vs SNR as $\rho$ is varied  for $20$ iterations of JED-ADMM and JED-AM in a $64 \times 80$ MIMO for independent Rayleigh channels.}
    \label{fig:JED_64X80_iid_varyrho_1}
\end{figure}

Fig. \ref{fig:JED_32X32_iid_varyrho_1} demonstrates that the choice of $\rho$ affects the BER performance, and a lower value of $\rho$ leads to an improvement in BER. For example, at $\text{SNR}=20$ dB, with $\rho = 4 \times \frac{\textstyle{\sigma^2_v}}{\textstyle{\sigma^2_h}}$, we get BER $\approx 10^{-2}$ while for $\rho = \frac{\textstyle{\sigma^2_v}}{\textstyle{\sigma^2_h}}$, the BER value drops to $\approx 2 \times 10^{-3}$. However, we still achieve an improvement over JED-AM irrespective of the chosen value of $\rho$. 
For $N>K$ in Fig. \ref{fig:JED_32X16_iid_varyrho_1}, we observe that for  $\rho = 4 \times \frac{\textstyle{\sigma^2_v}}{\textstyle{\sigma^2_h}}$, the performance of JED-ADMM and JED-AM are very close while for  $\rho = \frac{\textstyle{\sigma^2_v}}{\textstyle{\sigma^2_h}}$, JED-ADMM gives $\approx 0.5$ dB improvement in terms of $\text{SNR}$ at high $\text{SNR} > 10$ dB.
We also see from Fig. \ref{fig:JED_64X80_iid_varyrho_1} that for $N < K$, JED-ADMM with $\rho = \frac{\textstyle{\sigma^2_v}}{\textstyle{\sigma^2_h}}$ gives an order of magnitude lesser BER at $\text{SNR}=24$ dB than JED-AM. On the other hand, JED-ADMM with $\rho = 4 \times \frac{\textstyle{\sigma^2_v}}{\textstyle{\sigma^2_h}}$ performs similarly as JED-AM.

Thus, we observe that the choice of $\rho$ affects the BER performance of JED-ADMM which motivates us to consider letting the data decide the optimal value of $\rho$ in a trainable setting. This is the motivation for the unfolding of JED-ADMM iterations to give us the model-based deep learning JED-U-ADMM network.

\end{experiment}


\begin{experiment}{\textit{Study the effect of varying number of iterations, for a given $\rho$, in the presence of independent MIMO channels with $N=K$}}.
We study the BER performance of the proposed JED-ADMM algorithm as a function of the number of iterations for the case of independent and identically distributed (i.i.d) Rayleigh channels when the number of receive antennas equals the number of transmit antennas, $N = K$ for   $N = 16, 32$. The value of the ADMM penalty parameter $\rho = 4 \times \frac{\textstyle{\sigma^2_v}}{\textstyle{\sigma^2_h}}$ is kept constant throughout all the iterations.

\begin{figure*}[htb]
    \centering
    \subfloat[\label{fig:JED_16X16_iid_fixrho_varyiter}]{\includegraphics[width = 0.45\linewidth]{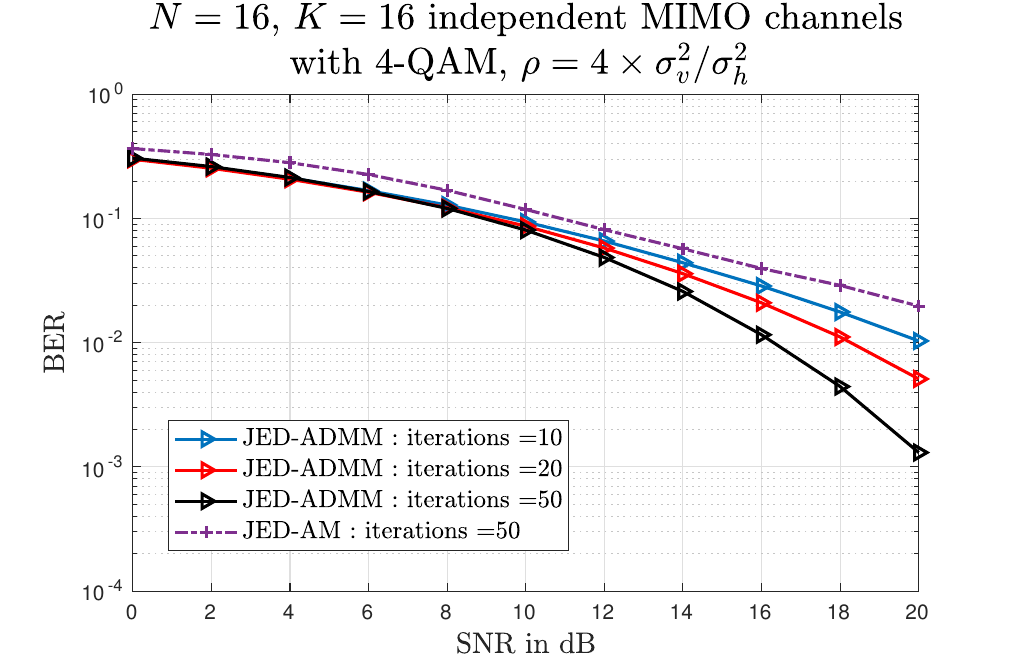}}
    \hfill
     \subfloat[\label{fig:JED_32X32_iid_fixrho_varyiter}]{\includegraphics[width = 0.45\linewidth]{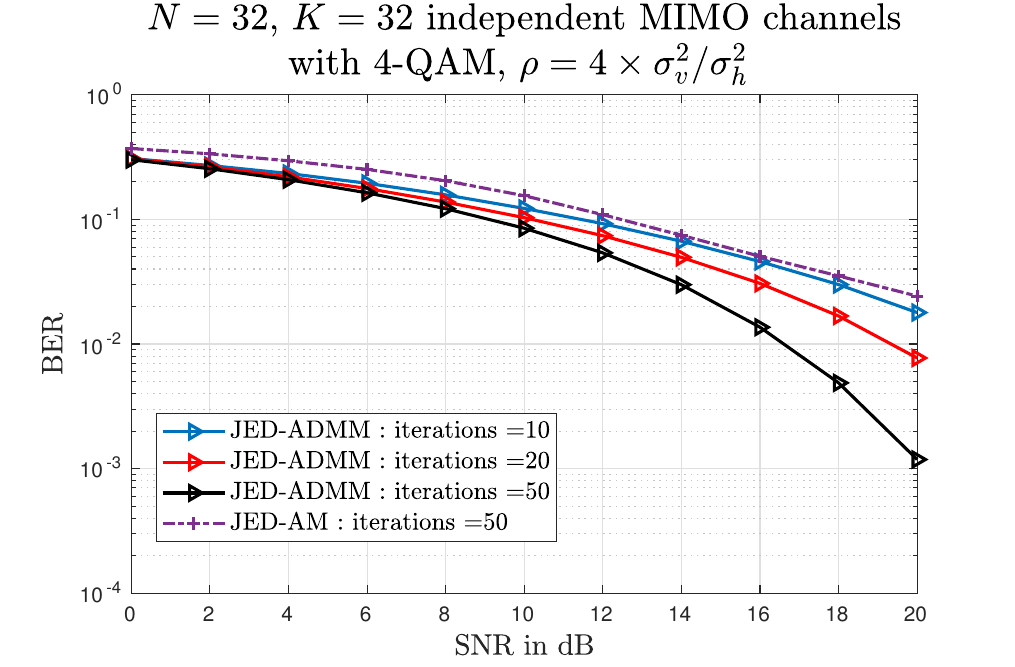}}
     
    \caption{ \footnotesize This plot shows the variation of BER vs $\text{SNR}$ depending on the number of iterations of JED-ADMM and JED-AM for (a) $16 \times 16$ and (b) $32 \times 32$ MIMO for independent Rayleigh channels.}
    \label{fig:varyIters_JED_fixrho}
\end{figure*}
In Fig. \ref{fig:varyIters_JED_fixrho}, we observe that with the increase in number of iterations, JED-ADMM yields a decreasing BER across all values of $\text{SNR}$. For example, in Fig. \ref{fig:varyIters_JED_fixrho} (a), at $\text{SNR} = 20$ dB, $50$ iterations of JED-AM gives us a BER of $\approx 10^{-2}$, whereas $50$ iterations of JED-ADMM produces BER of $\approx 10^{-3}$, which is an order of magnitude lower. We also note from Figs. \ref{fig:varyIters_JED_fixrho} (a) and \ref{fig:varyIters_JED_fixrho} (b) that for a given number of iterations and $\text{SNR}$, the performance is better for lower values of $N=K$.

It is to be noted, from Fig. \ref{fig:varyIters_JED_fixrho}, that even as low as $10$ iterations of JED-ADMM surpass $50$ iterations of JED-AM. Hence, even though the number of FLOPS per iteration is slightly higher for JED-ADMM than JED-AM (Table \ref{tab:detailed_FLOPScount_withmmsezf}), we see an overall reduction in complexity owing to the lesser number of iterations needed. Accordingly, for a $16\times16$ MIMO as depicted in Fig. \ref{fig:varyIters_JED_fixrho} (a), $50$ iterations of JED-AM requires $21299200$ FLOPS whereas $10$ iterations of JED-ADMM requires only $5570560$. Thus JED-ADMM reduces the overall FLOPS count by $\approx 75\%$ and yet achieves a lower BER.
\end{experiment}


\begin{experiment}{\textit{Study the effect of correlation coefficient $\rho_c$ at the receiver with $N=K$}}\\
We investigate how JED-ADMM performs for correlated channels,  that follow the Kronecker correlation channel model~\cite{correlationMIMOchannel} with the correlation coefficient $\rho_c = 0.5$ and $0.9$ as per \eqref{eq: corrAntMat}, for the $N = K$ MIMO configuration.   

\begin{figure*}[h]
    \centering
    \subfloat[\label{fig:JED_1May_16X16_corr}]{\includegraphics[width = 0.45\linewidth]{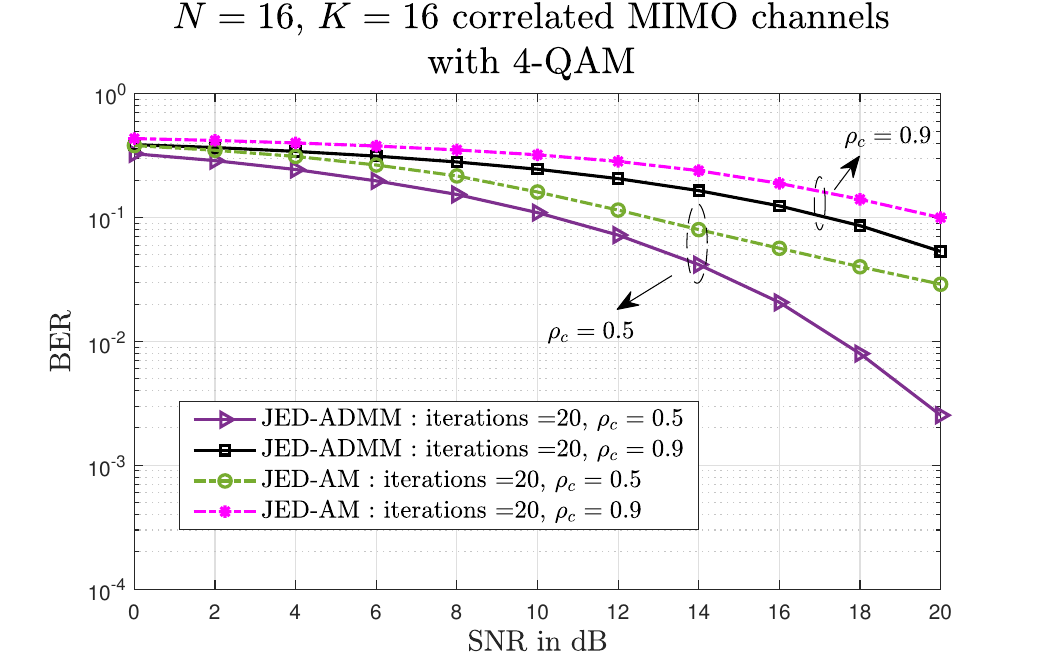}}
    \hfill
     \subfloat[\label{fig:JED_1May_32X32_corr}]{\includegraphics[width = 0.45\linewidth]{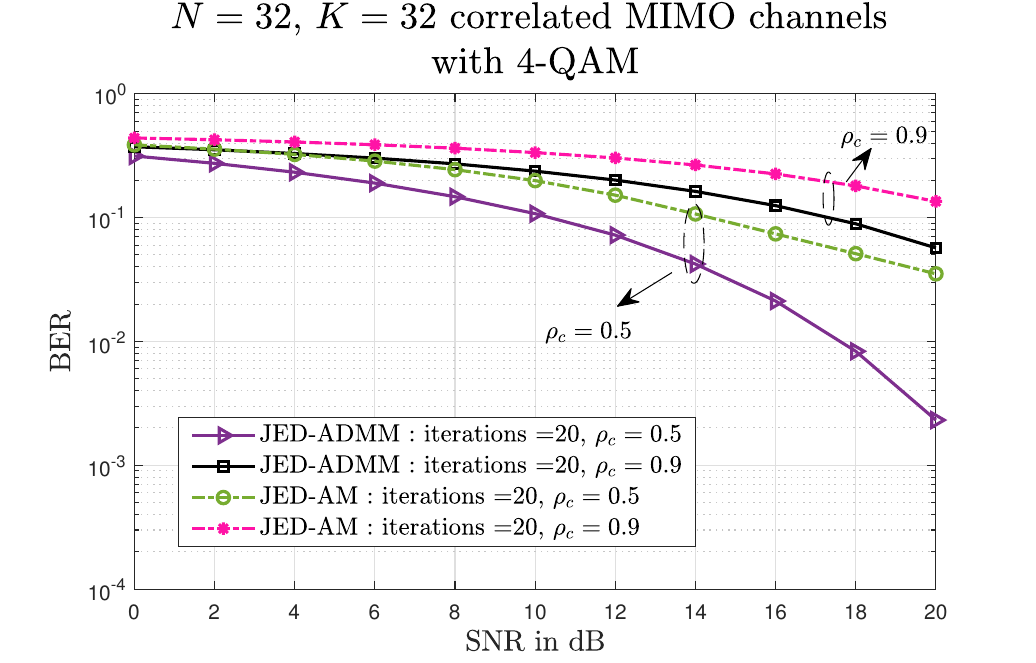}}
     
    \caption{ \footnotesize This plot shows the effect of the receiver correlation coefficient $\rho_c$ on the variation of BER vs $\text{SNR}$ for (a) $16 \times 16$ and (b) $32 \times 32$ MIMO for JED-ADMM and JED-AM for correlated channels~\cite{correlationMIMOchannel}. }
    \label{fig:JED_1May_corrChan}
\end{figure*}
We see from Fig. \ref{fig:JED_1May_corrChan}, as the correlation is decreased from $\rho_c = 0.9$ to $\rho_c = 0.5$, we see almost an order of magnitude lower BER for both JED-ADMM and JED-AM and that JED-ADMM performs better than JED-AM for both cases. For all values of $\rho_c$ considered and both the MIMO configurations of Figs. \ref{fig:JED_1May_corrChan} (a) and (b), JED-ADMM consistently shows better BER performance than JED-AM. Also, for a given \text{SNR} = $20$ dB and $N=K=16$, JED-ADMM achieves BER $\approx 10^{-3}$ whereas JED-AM achieves BER $\approx 10^{-2}$ i.e., an order of magnitude improvement in BER by JED-ADMM. It is to be noted from Figs. \ref{fig:JED_1May_corrChan} (a) and (b), that as the correlation coefficient $\rho_c$ decreases, the gap between the performance of JED-ADMM and JED-AM also increases. For $\rho_c=0.5$ and at an $\text{SNR}=20$ dB, JED-ADMM provides a considerably lesser BER, an order of magnitude lower, than JED-AM.

\end{experiment}


\subsubsection{Study the performance of JED-ADMM and proposed deep unfolded JED-U-ADMM}\label{sim_unfolded}
In the following experiments, we study our deep unfolded version JED-U-ADMM, in Algorithm \ref{alg: JED-U-ADMM}, and  JED-ADMM algorithm. 

\begin{experiment}{\textit{Study the effect of sharing parameters in the presence of independent MIMO channels with $N=K$.}}
We now study how using shared parameters $\left( \alpha, \theta, \gamma \right)$ across all the $L$ layers affects the BER performance of the unfolded network i.e. the \lq JED-U-ADMM : shared Params' architecture. We consider $10$ layers in a $16 \times 16$ MIMO system with i.i.d Rayleigh fading channel. Such a technique leads to having only $10+4 = 14$ trainable parameters instead of $40$ trainable parameters (if there was no sharing i.e. the \lq JED-U-ADMM : unshared Params' architecture).
\begin{figure}[h]
    \centering
    \includegraphics[width = 0.8\linewidth]{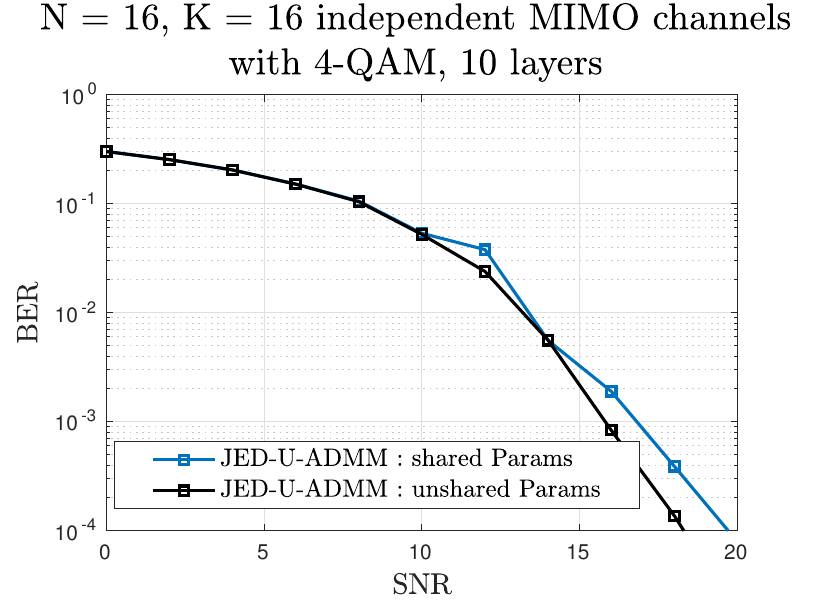}
    \caption{ \footnotesize This plot shows the effect of using shared parameters across the layers of the unfolded network for a $16 \times 16$ MIMO system for independent Rayleigh channels.}
    \label{fig:shared_unshared}
\end{figure}
It can be observed in Fig. \ref{fig:shared_unshared} that the loss in BER performance caused by the lesser number of trainable parameters in the \lq JED-U-ADMM : shared Params' architecture is negligible, even at higher $\text{SNR} > 16$ dB. The reduction in number of trainable parameters is also accompanied by a reduction in the computational complexity of the unfolded network during the training phase. Hence, to simplify the architecture, in the subsequent experiments, we use only the \lq JED-U-ADMM : shared Params' architecture.
\end{experiment}


\begin{experiment}{\textit{Study the effect of the number of layers on the unfolded network in the presence of independent MIMO channels with $N = K$.}}\\
In this experiment, we investigate how the BER performance of \lq JED-U-ADMM : shared Params' changes as the number of layers is varied from $5$ to $20$. We consider i.i.d Rayleigh fading channel in a $16 \times 16$ MIMO system.
\begin{figure}[h]
    \centering
    \includegraphics[width = 0.8\linewidth]{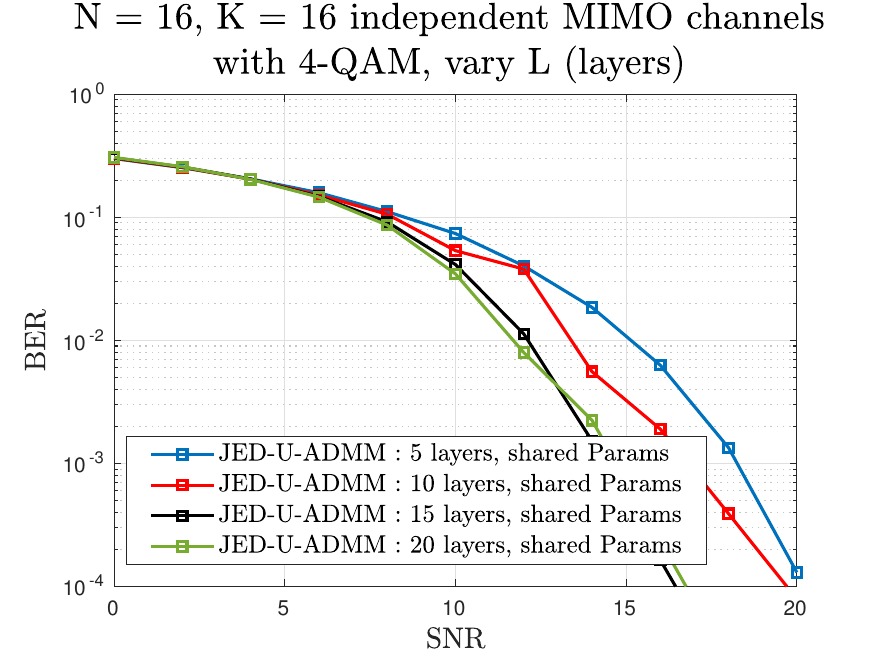}
    \caption{ \footnotesize This plot shows the effect of varying the number of layers in JED-U-ADMM in a $16 \times 16$ MIMO system for independent Rayleigh channels. }
    \label{fig:varyLayers_DLJEDADMM}
\end{figure}
In Fig ~\ref{fig:varyLayers_DLJEDADMM}, we observe that as the number of layers in increased from $5$ to $15$  the BER performance steadily improves. However, the network with $20$ layers provides marginal improvement and therefore, in the future set of experiments with \lq JED-U-ADMM : shared Params', we only consider the cases for $5$ and $10$ layers.
\end{experiment}


\begin{experiment}{\textit{Study the effect of unfolding in the presence of independent MIMO channels with $N=K$.}}\\
We study the BER performance of the unfolded algorithm JED-U-ADMM as a function of the number of layers of the neural network when the number of transmit and receive antennas are same in the presence of i.i.d Rayleigh fading channels.
\begin{figure*}[h]
    \centering
    \subfloat[\label{fig:14Apr_16X16_iid}]{\includegraphics[width = 0.45\linewidth]{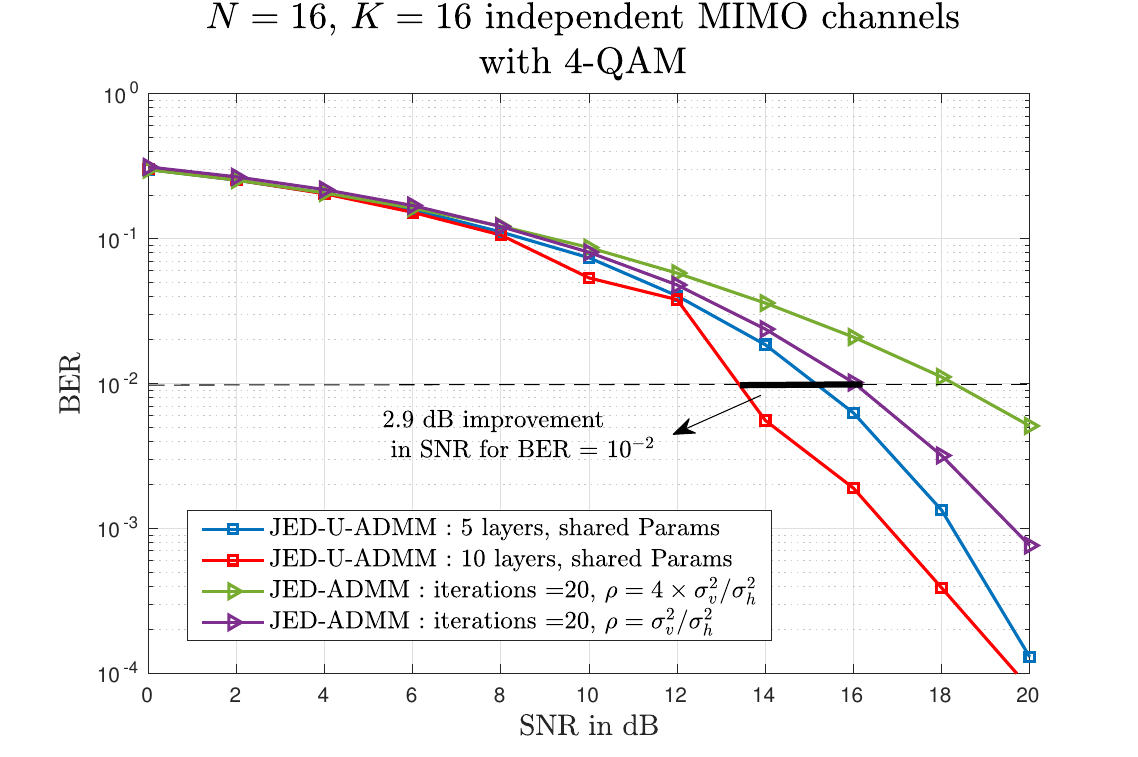}}
    \hfill
     \subfloat[\label{fig:14Apr_32X32_iid}]{\includegraphics[width = 0.45\linewidth]{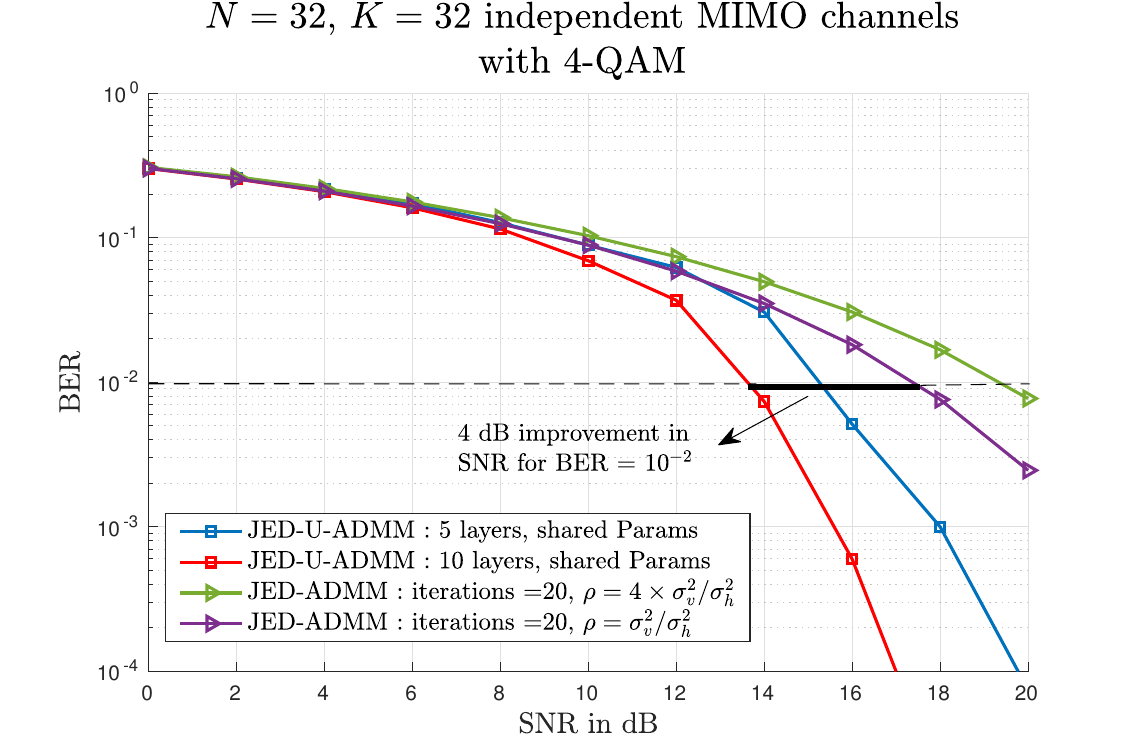}}
     
    \caption{ \footnotesize This plot shows the effect of deep unfolding on the variation of BER vs $\text{SNR}$ for (a) $16 \times 16$ and (b) $32 \times 32$ MIMO for independent Rayleigh channels.}
    \label{fig:14Apr_iidChan_NeqK}
\end{figure*}
Fig. \ref{fig:14Apr_iidChan_NeqK} demonstrates that unfolding leads to significant improvement in BER. 
    For $16 \times 16$ MIMO in Fig. \ref{fig:14Apr_iidChan_NeqK} (a), we observe that JED-U-ADMM with $10$ layers yields an $\text{SNR}$ improvement of $2.9$ dB in achieving a BER of $10^{-2}$, whereas for $32 \times 32$ MIMO in Fig. \ref{fig:14Apr_iidChan_NeqK} (b), we get an $\text{SNR}$ improvement of $4$ dB for a BER of $10^{-2}$.
    Comparing in terms of BER obtained at a given $\text{SNR}$, we see that for the $16 \times 16$ MIMO in Fig. \ref{fig:14Apr_iidChan_NeqK} (a), JED-U-ADMM needs $10$ layers to reach BER $\approx 10^{-3}$ at $\text{SNR}=16$ dB, which is an order of improvement over JED-ADMM that requires $20$ iterations to give a BER $\approx 10^{-2}$ at the same $\text{SNR}$. 
We also note that for $16 \times 16$ MIMO in Fig. \ref{fig:14Apr_iidChan_NeqK} (a), JED-U-ADMM with $5$ layers has almost similar performance as JED-ADMM with $20$ iterations and $\rho = \sigma^2_v/\sigma^2_h$. On the other hand, Fig. \ref{fig:14Apr_iidChan_NeqK} (b) shows that for $32 \times 32$ MIMO system, even $5$ layers of JED-U-ADMM shows more than $2$ dB $\text{SNR}$ improvement for SNR $ > 14$ dB. Thus, as we increase the system size, deep unfolding seems to perform much better than conventional iterative methods.

\end{experiment}


\begin{experiment}{\textit{Study the effect of varying the ratio $N/K$ in an unfolded network in the presence of independent MIMO channels.}}\\
This experiment showcases the performance of JED-U-ADMM when we operate in the standard MIMO regime of $N > K$ and in the overloaded MIMO regime of $N <K$.

\begin{figure*}[h]
    \centering
   
     \subfloat[\label{fig:14Apr_32X16_iid}]{\includegraphics[width = 0.45\linewidth]{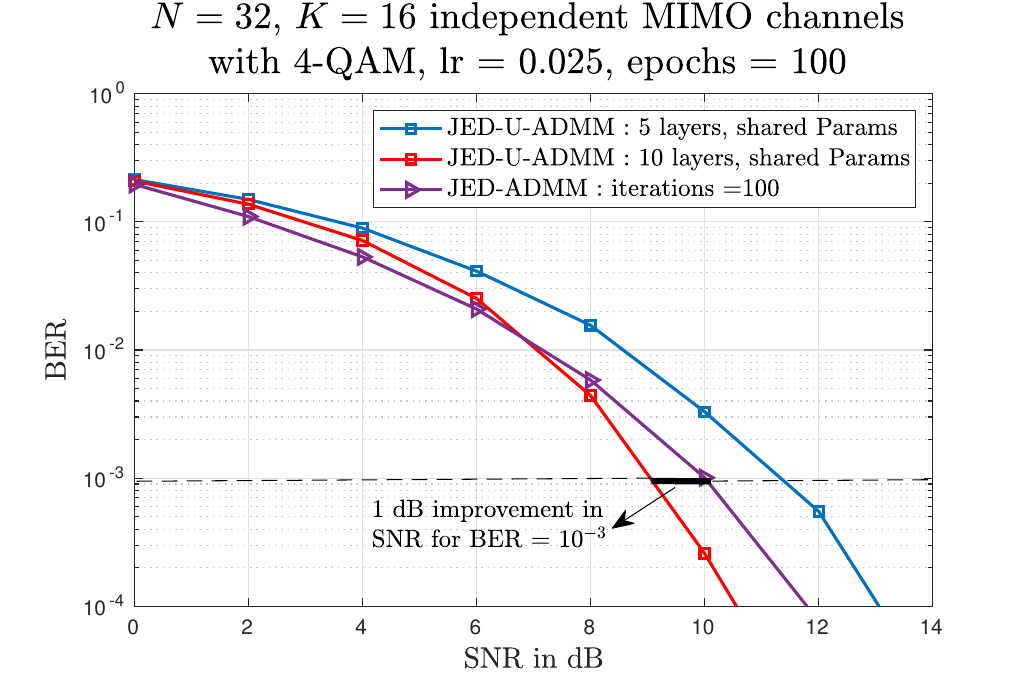}}
     \hfill
     \subfloat[\label{fig:14Apr_64X80_iid}]{\includegraphics[width = 0.45\linewidth]{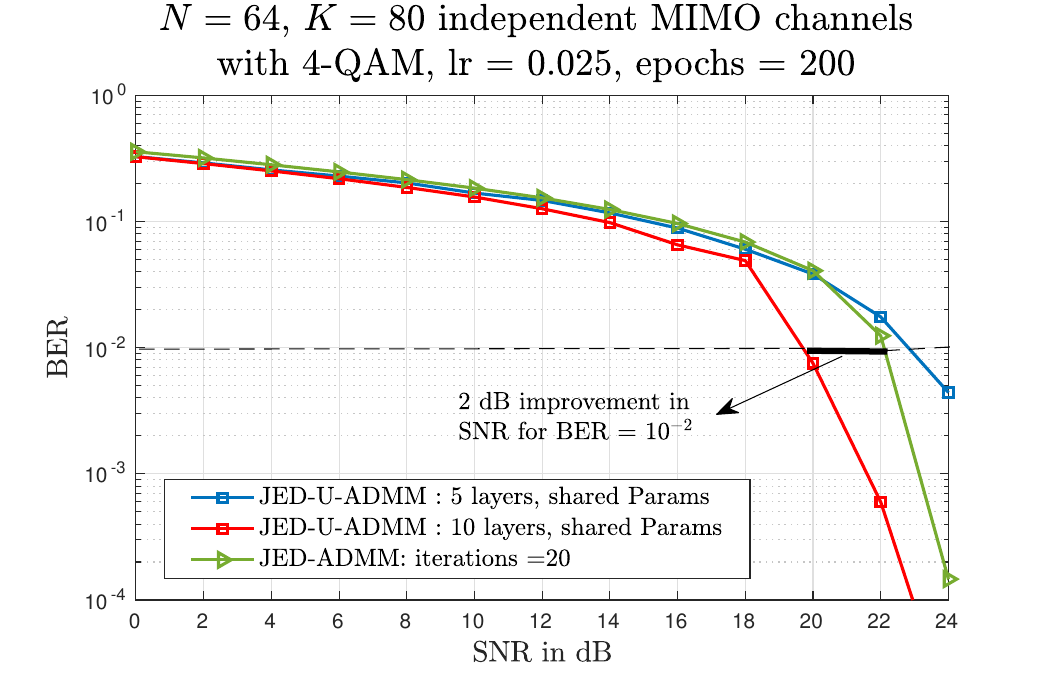}}
    \caption{ \footnotesize This plot shows the variation of BER vs $\text{SNR}$ for (a) $N > K$ : $32 \times 16$ MIMO and (b) $N < K$ : $64 \times 80$ for independent Rayleigh channels.}
    \label{fig:14Apr_iidChan}
\end{figure*}
For $N>K$ (the standard MIMO regime), we had observed from Fig. \ref{fig:JED_32X16_iid_varyrho_1} that JED-ADMM needs a large number of iterations of about $100$ to outperform JED-AM. However, with deep unfolding, we can improve the performance, especially at high $\text{SNR}$ values by about $1$ dB, as compared to JED-ADMM, with even $10$ layers as shown in Fig. \ref{fig:14Apr_iidChan} (a). 
\newline For the overloaded MIMO in Fig. \ref{fig:14Apr_iidChan} (b), we see that JED-U-ADMM needs only $10$ layers to consistently yield a lower BER than JED-ADMM. At $\text{SNR}$ = $20$ dB, JED-U-ADMM with $10$ layers gives a $2$ dB improvement in $\text{SNR}$ compared to $20$ iterations of JED-ADMM. As an example, at $\text{SNR}=22$ dB, we need $20$ iterations of JED-ADMM yield a BER of $10^{-2}$ whereas even $10$ layers of JED-U-ADMM gives us BER $< 10^{-3}$.

\end{experiment}


\begin{experiment}{\textit{Study the effect of correlation $\rho_c$ at the receiver antennas for unfolded network with $N\geq K$.}}
In this experiment, we demonstrate the performance of JED-ADMM and JED-U-ADMM, 
for the case of correlated channels. We consider the Kronecker correlation channel model~\cite{correlationMIMOchannel} with the correlation coefficient $\rho_c$ at the receiver. 

\begin{figure*}[h]
    \centering
    \subfloat[\label{fig:14Apr_16X16_corr}]{\includegraphics[width = 0.45\linewidth]{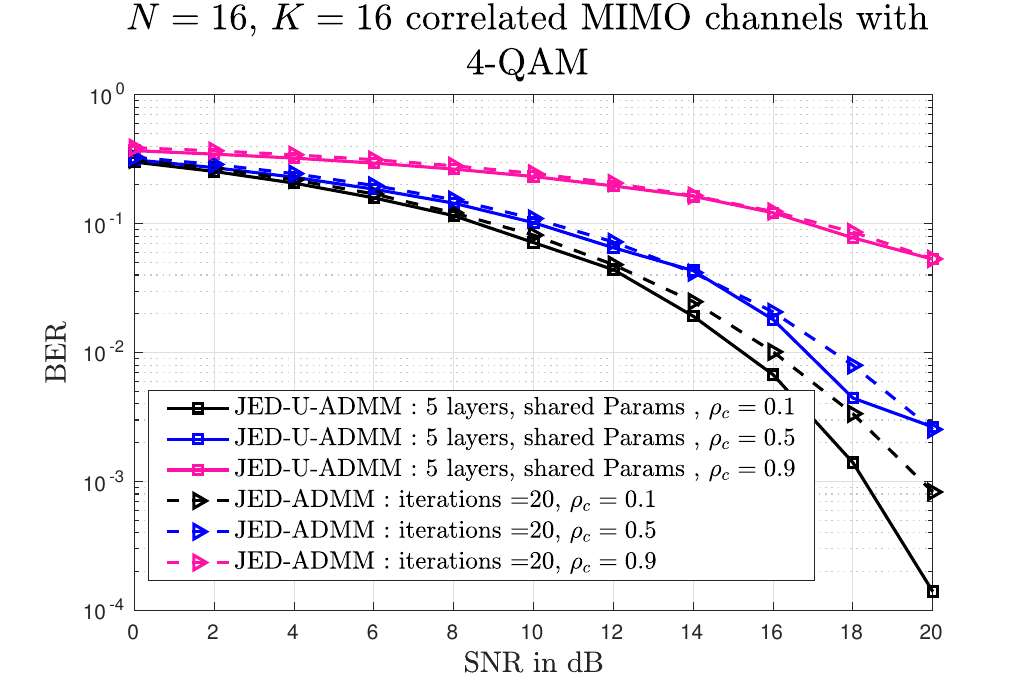}}
    \hfill
     \subfloat[\label{fig:14Apr_32X16_corr}]
     {\includegraphics[width = 0.45\linewidth]{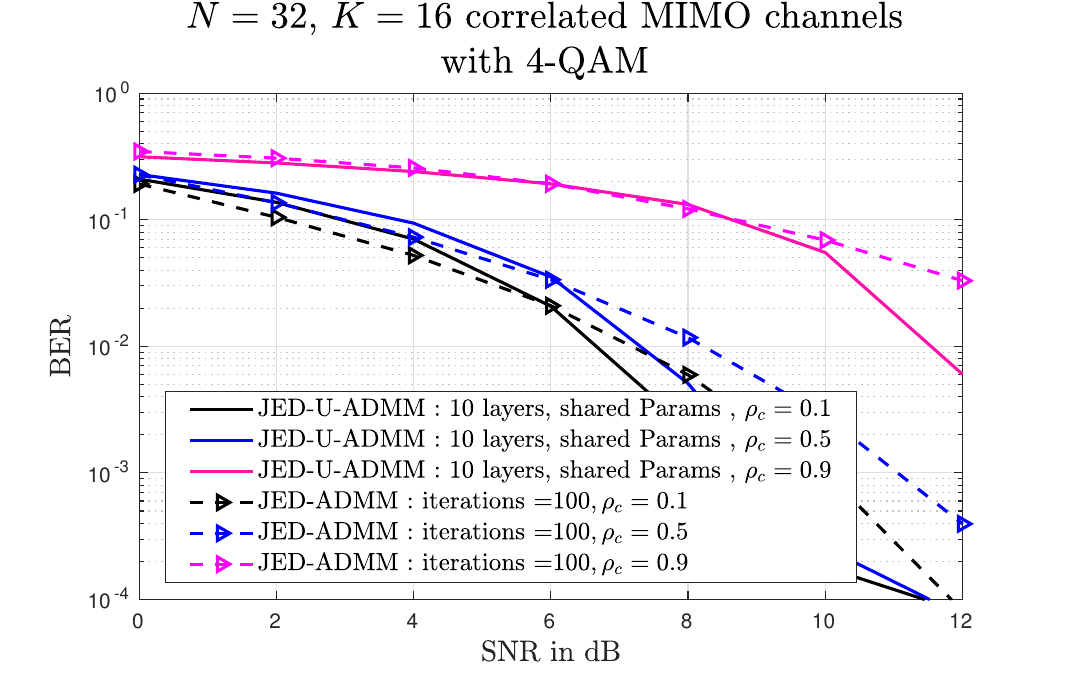}}
     
    \caption{ \footnotesize This plot shows the variation of BER vs $\text{SNR}$ for (a) $16 \times 16$ and (b) $32 \times 16$ MIMO where the channels are correlated at the Rx with $\rho_c$ \eqref{eq: corrAntMat} for correlated channels~\cite{correlationMIMOchannel}. }
    \label{fig:14Apr_corr}
\end{figure*}

We observe from Fig. \ref{fig:14Apr_corr} that as the correlation at the receiver antennas increases, both JED-ADMM and JED-U-ADMM degrade in performance. For the case of $16 \times 16$ MIMO, we compare $5$ layers of JED-U-ADMM with $20$ iterations of JED-ADMM in \ref{fig:14Apr_corr} (a). We see that for $\rho_c=0.1$, using unfolding results in slightly lesser BER of JED-U-ADMM than JED-ADMM, with the difference being more prominent at higher $\text{SNR}>16$ dB. However, for $\rho_c = 0.5$ and $0.9$, the BER curves of JED-U-ADMM and JED-ADMM are almost the same.
\newline In Fig. \ref{fig:14Apr_corr} (b), we present the BER curves for $32 \times 32$ with $10$ layers of JED-U-ADMM and $20$ iterations of JED-ADMM. For $\rho_c = 0.1$ and $0.5$, we see that unfolding improves the BER performance throughout the range of $\text{SNR}$ values presented, whereas for $\rho_c = 0.9$, the improvement is visible only for $\text{SNR} > 10$ dB.

\end{experiment}

\begin{experiment}{\textit{Study the effect of unfolding for a fixed $K=16$ and varying $N$ in the presence of independent MIMO channels.}}\\
We study the performance of the proposed algorithms by fixing the number of UEs $K=16$ and vary the number of BS antennas $N=8,16,32,48,64$. 

\begin{figure}[h]
    \centering
    \includegraphics[width = 0.8\linewidth]{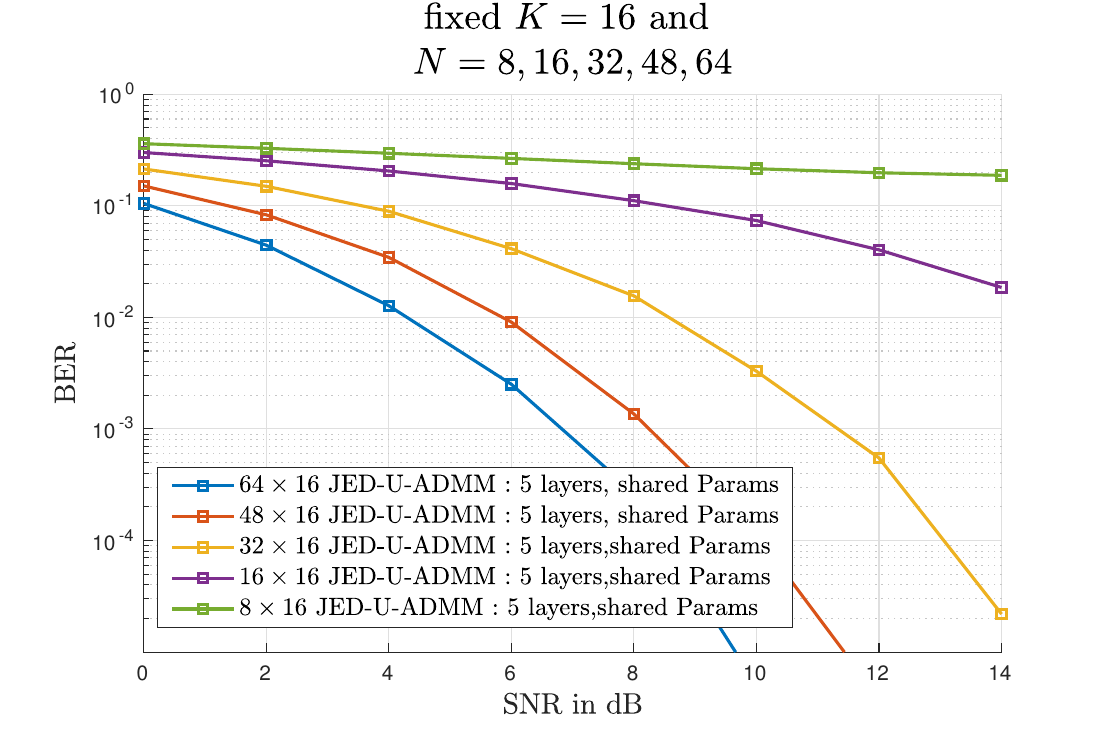}
    \caption{ \footnotesize This plot shows the variation of BER vs $\text{SNR}$ for $K=16$, $N=8,16,32,48,64$ MIMO for independent Rayleigh channels. }
    \label{fig:BERfixedK16_varyN}
\end{figure}

As $N$ increases, the system transitions from an overloaded scenario to the standard MIMO yielding a lower BER. As $N$ is increased from to $16$ to $32$, the BER falls drastically from $\approx 10^{-2}$ to $\approx 10^{-5}$ at $\text{SNR} = 14$ dB. Upon further increasing $N$ to $48$ and $64$, we note that the margin of improvement in performance gradually reduces.

\end{experiment}

We now present a summary of the observations in the next section.

\section{Discussion and Conclusion}
 In this paper, we propose two algorithms for Joint Channel Estimation and Symbol Detection, JED-ADMM and its unfolded version, JED-U-ADMM. The ADMM based algorithms for symbol detection exploit the non-smooth constraint arising from the QAM structure of the data symbols which significantly improve the BER performance compared to the state-of-the-art JED-AM.  We provide a summary our findings, as per the experiments conducted.
\par
Considering the fully loaded MIMO $N=K$, our proposed algorithm JED-ADMM, in Algorithm \ref{alg: JED-ADMM}, results in an $\text{SNR}$ improvement of about $3$ dB for BER = $10^{-2}$ for $\rho = \frac{\textstyle{\sigma^2_v}}{\textstyle{\sigma^2_h}}$ chosen, as shown in Fig. \ref{fig:JED_32X32_iid_varyrho_1}. The unfolded algorithm JED-U-ADMM, in Algorithm \ref{alg: JED-U-ADMM}, further provides $3$-$4$ dB $\text{SNR}$ improvement over JED-ADMM as shown in Fig. \ref{fig:14Apr_iidChan_NeqK} (a), thereby demonstrating the power of model-based neural network architectures. For the standard MIMO regime $N>K$, JED-U-ADMM needs only $10$ iterations to outperform $100$ iterations of JED-ADMM, by giving an $\text{SNR}$ improvement of $1$ dB whereas JED-ADMM requires $(\sim 100)$ iterations to yield an an $\text{SNR}$ improvement of $0.5$ dB over JED-AM, as depicted in Fig. \ref{fig:JED_32X16_iid_varyrho_1}. In the overloaded scenario $N < K$, we note from Fig. \ref{fig:JED_64X80_iid_varyrho_1} that JED-ADMM continues to exhibit waterfall decrease in BER whereas JED-AM saturates around BER of $2\times 10^{-2}$. Fig. \ref{fig:14Apr_iidChan} (b) depicts an additional $\text{SNR}$ improvement of $2$ dB by JED-U-ADMM. Furthermore, under correlated channel model
in Fig. \ref{fig:JED_1May_corrChan}, JED-ADMM markedly lowers the BER, especially for lower values of $\rho_c$. The unfolded network JED-U-ADMM further leads to slight improvement in BER performance over JED-ADMM in Fig. \ref{fig:14Apr_corr}.
\par Hence, we conclude that both of our proposed algorithms JED-ADMM and JED-U-ADMM enhance the BER performance in MIMO communication systems over the existing JED technique in \cite{Asilomar2019}. The deep unfolded network JED-U-ADMM provides us with an additional $\text{SNR}$ gain over the iterative JED-ADMM in much fewer iterations $\leq 10$, thus reducing the computational complexity. The unfolded network contains few trainable parameters and has simple update equations. In our future work, we plan to extend the proposed scheme to higher orders of modulation.


\bibliographystyle{IEEEtran}
\bibliography{doubleColumn_finalJournal_ver1}

\begin{thebibliography}{10}
\providecommand{\url}[1]{#1}
\csname url@samestyle\endcsname
\providecommand{\newblock}{\relax}
\providecommand{\bibinfo}[2]{#2}
\providecommand{\BIBentrySTDinterwordspacing}{\spaceskip=0pt\relax}
\providecommand{\BIBentryALTinterwordstretchfactor}{4}
\providecommand{\BIBentryALTinterwordspacing}{\spaceskip=\fontdimen2\font plus
\BIBentryALTinterwordstretchfactor\fontdimen3\font minus
  \fontdimen4\font\relax}
\providecommand{\BIBforeignlanguage}[2]{{%
\expandafter\ifx\csname l@#1\endcsname\relax
\typeout{** WARNING: IEEEtran.bst: No hyphenation pattern has been}%
\typeout{** loaded for the language `#1'. Using the pattern for}%
\typeout{** the default language instead.}%
\else
\language=\csname l@#1\endcsname
\fi
#2}}
\providecommand{\BIBdecl}{\relax}
\BIBdecl

\bibitem{bjornson2016massive}
E.~Björnson, E.~G. Larsson, and T.~L. Marzetta, ``Massive {MIMO}: ten myths
  and one critical question,'' \emph{IEEE Communications Magazine}, vol.~54,
  no.~2, pp. 114--123, 2016.

\bibitem{sah2016improved}
Abhishek, A.~K. Sah, and A.~K. Chaturvedi, ``Improved sparsity behaviour and
  error localization in detectors for large {MIMO} systems,'' in \emph{IEEE
  Globecom Workshops (GC Wkshps)}, 2016, pp. 1--6.

\bibitem{sparse_error_recovery}
E.~C. Marques, N.~Maciel, L.~Naviner, H.~Cai, and J.~Yang, ``A review of sparse
  recovery algorithms,'' \emph{IEEE access}, vol.~7, pp. 1300--1322, 2018.

\bibitem{ran2017sparse}
R.~Ran, J.~Wang, S.~K. Oh, and S.~N. Hong, ``Sparse-aware minimum mean square
  error detector for {MIMO} systems,'' \emph{IEEE Communications Letters},
  vol.~21, no.~10, pp. 2214--2217, 2017.

\bibitem{IW-SOAV}
R.~Hayakawa and K.~Hayashi, ``Convex optimization-based signal detection for
  massive overloaded {MIMO} systems,'' \emph{IEEE Transactions on Wireless
  Communications}, vol.~16, no.~11, pp. 7080--7091, 2017.

\bibitem{TPG}
S.~Takabe, M.~Imanishi, T.~Wadayama, and K.~Hayashi, ``Deep learning-aided
  projected gradient detector for massive overloaded mimo channels,'' in
  \emph{IEEE International Conference on Communications (ICC)}, 2019, pp. 1--6.

\bibitem{dlsd}
N.~T. Nguyen, K.~Lee, and H.~Dai, ``Application of deep learning to sphere
  decoding for large {MIMO} systems,'' \emph{IEEE Transactions on Wireless
  Communications}, vol.~20, no.~10, pp. 6787--6803, 2021.

\bibitem{albreem2021deep}
M.~A. Albreem, A.~H. Alhabbash, S.~Shahabuddin, and M.~Juntti, ``Deep learning
  for massive {MIMO} uplink detectors,'' \emph{IEEE Communications Surveys \&
  Tutorials}, vol.~24, no.~1, pp. 741--766, 2021.

\bibitem{jeon2015optimality}
C.~Jeon, R.~Ghods, A.~Maleki, and C.~Studer, ``Optimality of large {MIMO}
  detection via approximate message passing,'' in \emph{2015 IEEE International
  Symposium on Information Theory (ISIT)}.\hskip 1em plus 0.5em minus
  0.4em\relax IEEE, 2015, pp. 1227--1231.

\bibitem{luo2010semidefinite}
Z.-Q. Luo, W.-K. Ma, A.~M.-C. So, Y.~Ye, and S.~Zhang, ``Semidefinite
  relaxation of quadratic optimization problems,'' \emph{IEEE Signal Processing
  Magazine}, vol.~27, no.~3, pp. 20--34, 2010.

\bibitem{balatsoukas2019deep}
A.~Balatsoukas-Stimming and C.~Studer, ``Deep unfolding for communications
  systems: A survey and some new directions,'' in \emph{IEEE International
  Workshop on Signal Processing Systems (SiPS)}.\hskip 1em plus 0.5em minus
  0.4em\relax IEEE, 2019, pp. 266--271.

\bibitem{DetNet_SPAWC}
N.~Samuel, T.~Diskin, and A.~Wiesel, ``Deep {MIMO} detection,'' in \emph{IEEE
  18th International Workshop on Signal Processing Advances in Wireless
  Communications (SPAWC)}.\hskip 1em plus 0.5em minus 0.4em\relax IEEE, 2017,
  pp. 1--5.

\bibitem{oampnet2}
H.~He, C.-K. Wen, S.~Jin, and G.~Y. Li, ``Model-driven deep learning for {MIMO}
  detection,'' \emph{IEEE Transactions on Signal Processing}, vol.~68, pp.
  1702--1715, 2020.

\bibitem{blind_EP}
K.~Ghavami and M.~Naraghi-Pour, ``Blind channel estimation and symbol detection
  for multi-cell massive {MIMO} systems by expectation propagation,''
  \emph{IEEE Transactions on Wireless Communications}, vol.~17, no.~2, pp.
  943--954, 2018.

\bibitem{Pbigamp}
J.~Zhang, X.~Yuan, and Y.-J.~A. Zhang, ``Blind signal detection in massive
  {MIMO}: Exploiting the channel sparsity,'' \emph{IEEE Transactions on
  Communications}, vol.~66, no.~2, pp. 700--712, 2018.

\bibitem{SBCE_2020}
X.~Kuai, X.~Yuan, W.~Yan, H.~Liu, and Y.~J. Zhang, ``Double-sparsity
  learning-based channel-and-signal estimation in massive {MIMO} with
  generalized spatial modulation,'' \emph{IEEE Transactions on Communications},
  vol.~68, no.~5, pp. 2863--2877, 2020.

\bibitem{yan2019semi}
W.~Yan and X.~Yuan, ``Semi-blind channel-and-signal estimation for uplink
  massive {MIMO} with channel sparsity,'' \emph{IEEE Access}, vol.~7, pp.
  95\,008--95\,020, 2019.

\bibitem{Asilomar2019}
B.~B. Yilmaz and A.~T. Erdogan, ``Channel estimation for massive {MIMO}: A
  semiblind algorithm exploiting qam structure,'' in \emph{2019 53rd Asilomar
  Conference on Signals, Systems, and Computers}, 2019, pp. 2077--2081.

\bibitem{lecun_unfolding}
K.~Gregor and Y.~LeCun, ``Learning fast approximations of sparse coding,'' in
  \emph{Proceedings of the 27th International Conference on International
  Conference on Machine Learning}, 2010, p. 399–406.

\bibitem{dublid_journal}
Y.~Li, M.~Tofighi, J.~Geng, V.~Monga, and Y.~C. Eldar, ``Efficient and
  interpretable deep blind image deblurring via algorithm unrolling,''
  \emph{IEEE Transactions on Computational Imaging}, vol.~6, pp. 666--681,
  2020.

\bibitem{yang2020admm}
Y.~Yang, J.~Sun, H.~Li, and Z.~Xu, ``{ADMM}-{CSN}et: A deep learning approach
  for image compressive sensing,'' \emph{IEEE Transactions on Pattern Analysis
  and Machine Intelligence}, vol.~42, no.~3, pp. 521--538, 2020.

\bibitem{usrnet}
K.~Zhang, L.~Van~Gool, and R.~Timofte, ``Deep unfolding network for image
  super-resolution,'' in \emph{IEEE/CVF Conference on Computer Vision and
  Pattern Recognition (CVPR)}.\hskip 1em plus 0.5em minus 0.4em\relax IEEE,
  2020, pp. 3214--3223.

\bibitem{unrolling_review}
V.~Monga, Y.~Li, and Y.~C. Eldar, ``Algorithm unrolling: Interpretable,
  efficient deep learning for signal and image processing,'' \emph{IEEE Signal
  Processing Magazine}, vol.~38, no.~2, pp. 18--44, 2021.

\bibitem{softoutput_haochuan}
H.~Song, X.~You, C.~Zhang, and C.~Studer, ``Soft-output joint channel
  estimation and data detection using deep unfolding,'' in \emph{2021 IEEE
  Information Theory Workshop (ITW)}, 2021, pp. 1--5.

\bibitem{mmnet}
M.~Khani, M.~Alizadeh, J.~Hoydis, and P.~Fleming, ``Adaptive neural signal
  detection for massive {MIMO},'' \emph{IEEE Transactions on Wireless
  Communications}, vol.~19, no.~8, pp. 5635--5648, 2020.

\bibitem{oampnet}
H.~He, C.-K. Wen, S.~Jin, and G.~Y. Li, ``A model-driven deep learning network
  for {MIMO} detection,'' in \emph{IEEE Global Conference on Signal and
  Information Processing (GlobalSIP)}.\hskip 1em plus 0.5em minus 0.4em\relax
  IEEE, 2018, pp. 584--588.

\bibitem{lordnet}
S.~Khobahi, N.~Shlezinger, M.~Soltanalian, and Y.~C. Eldar, ``Lord-net:
  Unfolded deep detection network with low-resolution receivers,'' \emph{IEEE
  Transactions on Signal Processing}, vol.~69, pp. 5651--5664, 2021.

\bibitem{jammer_studer}
G.~Marti and C.~Studer, ``Mitigating smart jammers in {MU}-{MIMO} via joint
  channel estimation and data detection,'' in \emph{ICC 2022 - IEEE
  International Conference on Communications}, 2022, pp. 1336--1342.

\bibitem{jed_cellfree}
H.~Song, T.~Goldstein, X.~You, C.~Zhang, O.~Tirkkonen, and C.~Studer, ``Joint
  channel estimation and data detection in cell-free massive {MU}- {MIMO}
  systems,'' \emph{IEEE Transactions on Wireless Communications}, vol.~21,
  no.~6, pp. 4068--4084, 2022.

\bibitem{unfolding_chanest_beamform}
K.~Kang, Q.~Hu, Y.~Cai, G.~Yu, J.~Hoydis, and Y.~C. Eldar, ``Mixed-timescale
  deep-unfolding for joint channel estimation and hybrid beamforming,''
  \emph{IEEE Journal on Selected Areas in Communications}, vol.~40, no.~9, pp.
  2510--2528, 2022.

\bibitem{unfolding_mixedadc}
L.~Xu, F.~Gao, T.~Zhou, S.~Ma, and W.~Zhang, ``Joint channel estimation and
  mixed-{ADC}s allocation for massive {MIMO} via deep learning,'' \emph{IEEE
  Transactions on Wireless Communications}, vol.~22, no.~2, pp. 1029--1043,
  2023.

\bibitem{admm_hnet}
M.-W. Un, M.~Shao, W.-K. Ma, and P.~Ching, ``Deep {MIMO} detection using {ADMM}
  unfolding,'' in \emph{2019 IEEE Data Science Workshop (DSW)}.\hskip 1em plus
  0.5em minus 0.4em\relax IEEE, 2019, pp. 333--337.

\bibitem{parna_icc}
P.~Sabeti, A.~Farhang, I.~Macaluso, N.~Marchetti, and L.~Doyle, ``Blind channel
  estimation for massive {MIMO}: A deep learning assisted approach,'' in
  \emph{IEEE International Conference on Communications (ICC)}, 2020, pp. 1--6.

\bibitem{GEM_tsp2022}
Y.~Zhang, J.~Sun, J.~Xue, G.~Y. Li, and Z.~Xu, ``Deep expectation-maximization
  for joint mimo channel estimation and signal detection,'' \emph{IEEE
  Transactions on Signal Processing}, vol.~70, pp. 4483--4497, 2022.

\bibitem{boyd2011distributed}
S.~Boyd, N.~Parikh, and E.~Chu, \emph{Distributed optimization and statistical
  learning via the alternating direction method of multipliers}.\hskip 1em plus
  0.5em minus 0.4em\relax Now Publishers Inc, 2011.

\bibitem{ADMIN_Studer}
S.~Shahabuddin, I.~Hautala, M.~Juntti, and C.~Studer, ``Admm-based
  infinity-norm detection for massive mimo: Algorithm and vlsi architecture,''
  \emph{IEEE Transactions on Very Large Scale Integration (VLSI) Systems},
  vol.~29, no.~4, pp. 747--759, 2021.

\bibitem{DUBLID}
Y.~Li, M.~Tofighi, V.~Monga, and Y.~C. Eldar, ``An algorithm unrolling approach
  to deep image deblurring,'' in \emph{IEEE International Conference on
  Acoustics, Speech and Signal Processing (ICASSP)}.\hskip 1em plus 0.5em minus
  0.4em\relax IEEE, 2019, pp. 7675--7679.

\bibitem{jed_ssp}
S.~Bhattacharya, K.~V.~S. Hari, and Y.~C. Eldar, ``Joint channel estimation and
  symbol detection in overloaded {MIMO} using {ADMM},'' in \emph{2023 IEEE
  Statistical Signal Processing Workshop (SSP)}, accepted for publication.

\bibitem{paszke2019pytorch}
A.~Paszke, S.~Gross, F.~Massa, A.~Lerer, J.~Bradbury, G.~Chanan, T.~Killeen,
  Z.~Lin, N.~Gimelshein, L.~Antiga \emph{et~al.}, ``Pytorch: An imperative
  style, high-performance deep learning library,'' \emph{Advances in neural
  information processing systems}, vol.~32, 2019.

\bibitem{kingma2014adam}
D.~P. Kingma and J.~Ba, ``Adam: A method for stochastic optimization,''
  \emph{arXiv preprint arXiv:1412.6980}, 2014.

\bibitem{correlationMIMOchannel}
C.~Oestges, ``Validity of the kronecker model for {MIMO} correlated channels,''
  in \emph{IEEE 63rd Vehicular Technology Conference}, vol.~6.\hskip 1em plus
  0.5em minus 0.4em\relax IEEE, 2006, pp. 2818--2822.

\end{thebibliography}
\end{document}